\documentclass[twocolumn]{aastex631}
\def\msol{$M_{\odot}$\xspace}
\def\mstar{$M_{\star}$\xspace}
\def\mdot{$M_{\odot} \mbox{ yr}^{-1}$}
\def\lum{erg~s$^{-1}$}

\def\chandra{{\itshape Chandra\/}}

\def\hst{{\itshape HST\/}}
\def\jwst{{\itshape JWST\/}}

\def\xray{\hbox{X-ray}}

\def\etal{{et\,al.}}

\def\ltsima{$\; \buildrel < \over \sim \;$}
\def\simlt{\lower.5ex\hbox{\ltsima}}
\def\gtsima{$\; \buildrel > \over \sim \;$}
\def\simgt{\lower.5ex\hbox{\gtsima}}
\def\kms{\ifmmode{~{\rm km~s^{-1}}}\else{~km s$^{-1}$}\fi}
\def\lsim{\lower0.3em\hbox{$\,\buildrel <\over\sim\,$}}
\def\gsim{\lower0.3em\hbox{$\,\buildrel >\over\sim\,$}}
\def\msol{$M_\odot$}

\def\lum{erg~s$^{-1}$}

\def\sfr{$M_{\odot}$~yr$^{-1}$}

\def\aap{A\&A}

\def\apjl{ApJL}
\def\apjs{ApJS}

\def\nsrc{765}

\makeatletter
\DeclareRobustCommand{\HII}{%
  \mbox{H\check@mathfonts\fontsize\sf@size\z@\selectfont II}%
}
\makeatother

\usepackage{underscore}
\usepackage{longtable}
\usepackage{amsmath}
\usepackage{comment}
\usepackage{xspace}
\usepackage{booktabs}
\usepackage{hyperref}

\begin{document}


\title{{\large Constructing X-ray Spectral Models of Galaxies: Varying Contributions from X-ray Binary Populations with Host Galaxy Properties}}

\correspondingauthor{Emerson Gehr}
\email{egehr@uark.edu}

\author[0009-0008-4203-6112]{Emerson Gehr}
\affiliation{Department of Physics, University of Arkansas, 226 Physics Building, 825 West Dickson Street, Fayetteville, AR 72701, USA}

\author[0000-0003-2192-3296]{Bret~D.~Lehmer}
\affiliation{Department of Physics, University of Arkansas, 226 Physics Building, 825 West Dickson Street, Fayetteville, AR 72701, USA}
\affiliation{Arkansas Center for Space and Planetary Sciences, University of Arkansas, 332 N. Arkansas Avenue, Fayetteville, AR 72701, USA}

\author[0000-0001-8525-4920]{Antara R. Basu-Zych}
\affiliation{Department of Physics, University of Maryland Baltimore County, Baltimore, MD 21250, USA}
\affiliation{NASA Goddard Space Flight Center, Code 662, Greenbelt, MD 20771, USA}
\affiliation{Center for Research and Exploration in Space Science and Technology, NASA/GSFC, Greenbelt, MD 20771, USA}

\author[0000-0003-3684-964X]{Konstantinos Kovlakas}
\affiliation{Institute of Space Sciences (ICE), CSIC, Campus UAB, Carrer de Can Magrans s/n, E-08193, Barcelona, Spain}
\affiliation{Institut d’Estudis Espacials de Catalunya (IEEC), Edifici RDIT, Campus UPC, E-08860 Castelldefels (Barcelona), Spain}

\author[0000-0002-7716-6223]{Vianney Lebouteiller}
\affiliation{Université Paris-Saclay, Université Paris Cité, CEA, CNRS, AIM, 91191, Gif-sur-Yvette, France}

\author[0000-0001-8473-5140]{Erik B. Monson}
\affiliation{Department of Physics and Astronomy, Middle Tennessee State University, 1301 E. Main Street Box 71, Murfreesboro, TN 37132, USA}

\author[0009-0002-9967-3557]{Izabela Pavel}
\affiliation{Department of Physics and Astronomy, University of New Mexico, 210 Yale Blvd NE, Albuquerque, NM 87016, USA}
\affiliation{Postgraduate Program in High Energy Physics, Astrophysics and Cosmology, Facultat de Ciències, Campus UAB, 08193 Bellaterra (Barcelona), Spain}
\affiliation{Institute of Space Sciences (ICE), CSIC, Campus UAB, Carrer de Can Magrans s/n, E-08193, Barcelona, Spain}
\affiliation{Department of Physics, University of Arkansas, 226 Physics Building, 825 West Dickson Street, Fayetteville, AR 72701, USA}

\author[0000-0002-3703-0719]{Chris Richardson}
\affiliation{Department of Physics \& Astronomy, Elon University, 100 Campus Drive, Elon, NC 27244, USA}

\author[0009-0004-6699-8341]{Maxime Varese}
\affiliation{Université Paris-Saclay, Université Paris Cité, CEA, CNRS, AIM, 91191, Gif-sur-Yvette, France}

\author[0009-0002-3429-9103]{Zachary Wilson}
\affiliation{Department of Physics, University of Arkansas, 226 Physics Building, 825 West Dickson Street, Fayetteville, AR 72701, USA}





\begin{abstract}

Recent work has shown that the emission from X-ray binary (XRB) populations in galaxies varies with stellar mass (\mstar), star formation rate (SFR), and metallicity ($Z$). Such scaling relations are widely used to predict the XRB contributions to galaxy-integrated X-ray luminosities including studies focused on dwarf active galactic nuclei (AGN) and the X-ray radiation field during the epoch of heating in the early ($z \geq 8$) universe. However, as galaxies approach low SFR and low $Z$, the relatively shallow slope of the XRB luminosity function (XLF) can yield very large stochastic variations in the total X-ray luminosity expected from the XRB population, for fixed values of \mstar, SFR, and $Z$. We have created a procedure to statistically sample any XLF and model total X-ray spectra for XRB populations and their stochastic uncertainties. We demonstrate the accuracy of this procedure using data for galaxies ranging from high to low \mstar, SFR, and $Z$ and generating X-ray spectral models consistent with \chandra\ observations. For galaxies that lie on the galactic main-sequence, we can relate SFR and $Z$ to \mstar using established \mstar-SFR and \mstar-$Z$ relations. Applying these relations, we construct main-sequence (MS) XRB spectral models, which provide typical XRB spectral shapes, normalizations, and uncertainties as a function of \mstar. The spectral model library associated with this work is available at \url{https://doi.org/10.5281/zenodo.20126734}.

\end{abstract}





\section{Introduction and Motivation}\label{sec:intro}
 
X-ray binaries (XRBs) are binary star systems that consist of a compact object (CO), typically a black hole (BH) or neutron star (NS), that accretes from a companion donor star. XRBs produce an extraordinary amount of powerful X-ray emission that has implications for the evolution of their host galaxy. The companion star transfers material to the compact object typically through either Roche lobe overflow or stellar winds, which accrete onto the compact object forming a hot X-ray emitting accretion disk. However, not all XRBs are formed through accretion physics. A subclass of XRBs, Be XRBs, are formed through decretion disks and may also be a significant population in a galaxy, both in numbers and in luminosity \citep[e.g.,][]{rocha2024or}. The nature of the compact object, formation mechanism, and the mass transfer rate determine spectral characteristics \citep[e.g.,][]{done_modelling_2007,weng2024x}. XRBs are typically divided into two main types: high-mass X-ray binaries (HMXBs) and low-mass X-ray binaries (LMXBs) and their classification depends on the mass of the system’s companion star. More massive stars are known to have shorter lifetimes, meaning HMXBs are more commonly found in galaxies that are more active in star formation, such as spiral and dwarf galaxies \citep[e.g.,][]{grimm_high-mass_2003,geda_high-mass_2024}. LMXBs are typically found in older populations and elliptical galaxies \citep[e.g.,][]{boroson_revisiting_2011,zhang_dependence_2012}. 

Empirical studies using samples of galaxies spanning $z \approx$~0--3 have shown that the X-ray power output from XRB populations strongly varies with stellar mass (\mstar), star formation rate (SFR), and metallicity ($Z$) \citep[see, e.g.,][for a recent review]{gilfanov_2022}. Specifically, the total luminosity of HMXBs scales with SFR and $Z$ \citep[e.g.,][]{grimm_x-ray_2003, mineo_x-ray_2012,brorby_enhanced_2016,lehmer_metallicity_2021}, while the total luminosity of LMXBs scales with \mstar \citep[e.g.,][]{gilfanov_statistical_2004, zhang_dependence_2012,kouroumpatzakis_sub-galactic_2020, lehmer_x-ray_2020}. Such scaling relations are widely used in a variety of applications, including, e.g., testing binary population synthesis models \citep[e.g.,][]{fragos_x-ray_2013,fragos_energy_2013,Misra2023,bray_2025}; estimating supermassive black hole (SMBH) occupation fractions among galaxies \citep[e.g.,][]{mezcua_2018,burke_2025,zou_2025}; and modeling the contributions of X-ray heating of the $z \simgt 8$ intergalactic medium \citep[e.g.,][]{mesinger_2013,pacucci_2014,madau_fragos2017,kovlakas2022ionizing, nikolic_2024}.
Some other recent applications involving these scaling relations are spectral modeling of XRB populations hosting ultraluminous \xray\ sources (ULXs) \citep[e.g.,][]{garofali2024modeling}, which is often used for interpreting JWST observations \citep[e.g.,][]{mingozzi2025exploring} and connecting these ULXs to HeII emission \citep[e.g.,][]{senchyna2020high, simmonds2021can}. These applications often assume slopes and shapes of spectral models, but binary-star processes must be explicitly accounted for in order to accurately represent spectral contributions. Typically, fixed scaling relations are used to build spectral models, without accounting for a specific galaxy's characteristics and degree of stochasticity \citep[e.g.,][]{ranalli20032, colbert_old_2004, persic_galactic_2007, mineo_x-ray_2012, mineo2012x, aird_x-rays_2017, lehmer_x-ray_2019, lehmer_x-ray_2020}. We must consider the \mstar, SFR, and $Z$ of the galaxy and stochasticity of the X-ray luminosity function (XLF) to best represent the nature and behavior of their X-ray spectral models \citep[][]{gilfanov_statistical_2004}.   

In this paper, we provide a procedure for translating the \mstar, SFR, and $Z$ of galaxies into accurate X-ray spectral models of their underlying XRB populations, not including active galactic nuclei (AGN), in the X-ray band ($0.5-8$ keV). Our models are constructed using the combination of empirical scaling relations of XLFs with galaxy properties and the effects of stochastic scatter on population-integrated X-ray point-source spectra, and thus provides X-ray spectral models and their uncertainties that vary with \mstar, SFR, and $Z$. X-ray spectral models are critical in learning about the intrinsic properties of unresolved stellar populations that impact galaxy evolution \citep[e.g.,][]{madau_cosmic_2014}. 


In $\S$\ref{sec:xlf}, we present our adopted XLF scaling relations and discuss their key features across luminosity regimes.  In $\S$\ref{sec:statistics}, we describe our methods for sampling the XLF in detail and discuss the effects of intrinsic stochastic scatter on SFR and $Z$, and how this translates to estimating a population's total X-ray luminosity ($L_{\rm{X}}$). $\S$\ref{sec:models} outlines our procedure for modeling the integrated X-ray source population spectra. $\S$\ref{sec:results} tests our spectral model predictions against X-ray data for a sample of nearby resolved galaxies, and provides models of X-ray spectra across the galaxy main sequence (MS). Finally, in $\S$\ref{sec:sum}, we summarize key results and explore future applications of this work.

\section{Adopted XLF Scaling Relations}\label{sec:xlf}


We begin our X-ray spectral modeling procedures by selecting XLF scaling relations that allow us to map galaxy properties to XRB population distributions.  We chose to adopt the SFR and $Z$ dependent HMXB scaling relations from \citet{lehmer_metallicity_2021} and the $M_\star$ dependent LMXB scaling relation from \citet{lehmer_x-ray_2019}.
The XLF scaling relations are specified below as equations~\ref{eq:HMXLF} and \ref{eq:LMXLF}, with adopted values of constant terms defined in Table~\ref{tab:const}:

\begin{table}
\tablenum{1}
\label{tab:const}
\begin{center}
\caption{Assumed XLF Model Parameters}
\begin{tabular}{lcr}
\hline\hline
\multicolumn{1}{l}{\sc Parameter Name} & {\sc Units} & \multicolumn{1}{c}{\sc Value} \\
\hline\hline
\multicolumn{3}{c}{HMXB Parameters (Eqn.~\ref{eq:HMXLF})}\\
\hline
$A_{\rm HM}$ \dotfill & (\sfr)$^{-1}$ & 1.29 \\
$\gamma_1$ \dotfill & & 1.74 \\
$\log L_{b, {\rm HM}}$\dotfill  & $\log$~(\lum)  & 38.54 \\
$\gamma_{2,\odot}$\dotfill  & &  1.16 \\
$\log L_{c,\odot}$\dotfill  & $\log$~(\lum)  & 39.98 \\
$\dfrac{d \gamma_2}{d \log Z}$\dotfill  & dex$^{-1}$  & 1.34 \\
$\dfrac{d\log L_{c,\rm HM}}{d \log Z}$\dotfill  & dex~dex$^{-1}$  & 0.60 \\
\hline
\multicolumn{3}{c}{LMXB Parameters (Eqn.~\ref{eq:LMXLF})}\\
\hline
$K_{\rm LM}$ \dotfill & ($10^{11}$ \msol)$^{-1}$ & 26.0 \\
$\alpha_1$ \dotfill &   & 1.31 \\
$\log L_{b, {\rm LM}}$\dotfill  & $\log$~(\lum)  & 38.3 \\
$\alpha_2$ \dotfill &  & 2.57 \\
$\log L_{c, {\rm LM}}$\dotfill  & $\log$~(\lum)  & 40.8 \\
\hline
\end{tabular}
\end{center}
Note - The above parameter values correspond to terms in Equations~\ref{eq:HMXLF} and \ref{eq:LMXLF}.
\end{table}

\begin{equation}
    \begin{split}
        \left.\frac{dN ({\rm SFR},Z)}{dL}\right \vert_{\rm HM}= {\rm SFR}\, A_{\rm HM}\, \exp[-L/L_{c, {\rm HM}}(Z)] \\ \times 
        \begin{cases}
            L^{-\gamma_1}, & (L < L_{b, {\rm HM}})\\
            L_{b, {\rm HM}}^{\gamma_2(Z)-\gamma_1}L^{-\gamma_2(Z)}, & (L > L_{b, {\rm HM}})
        \end{cases}
    \end{split}
    \label{eq:HMXLF}
\end{equation}
where
$$\gamma_2(Z) = \gamma_{\rm 2,\odot} + \frac{d \gamma_2}{d \log Z}\log (Z/Z_\odot),$$
$$\log L_{c, {\rm HM}}(Z) = \log L_{c, \odot} + \frac{d \log L_{c}}{d \log
Z}\log (Z/Z_\odot),$$
and
\begin{equation}
    \begin{split}
        \left.\frac{dN (M_\star)}{dL}\right \vert_{\rm LM} &\ = M_\star \, K_{\rm LM} \\ \times 
        &\ \begin{cases}
            L^{-\alpha_1}, & (L < L_{b, {\rm LM}})\\
            L_{b, {\rm LM}}^{\alpha_2-\alpha_1}L^{-\alpha_2}, & (L_{b, {\rm LM}} \leq L < L_{c, \rm LM})\\
            0, & (L \geq L_{c, {\rm LM}}).
        \end{cases}
    \end{split}
    \label{eq:LMXLF}
\end{equation}

\begin{figure*}
\centering
\includegraphics[width=\textwidth]{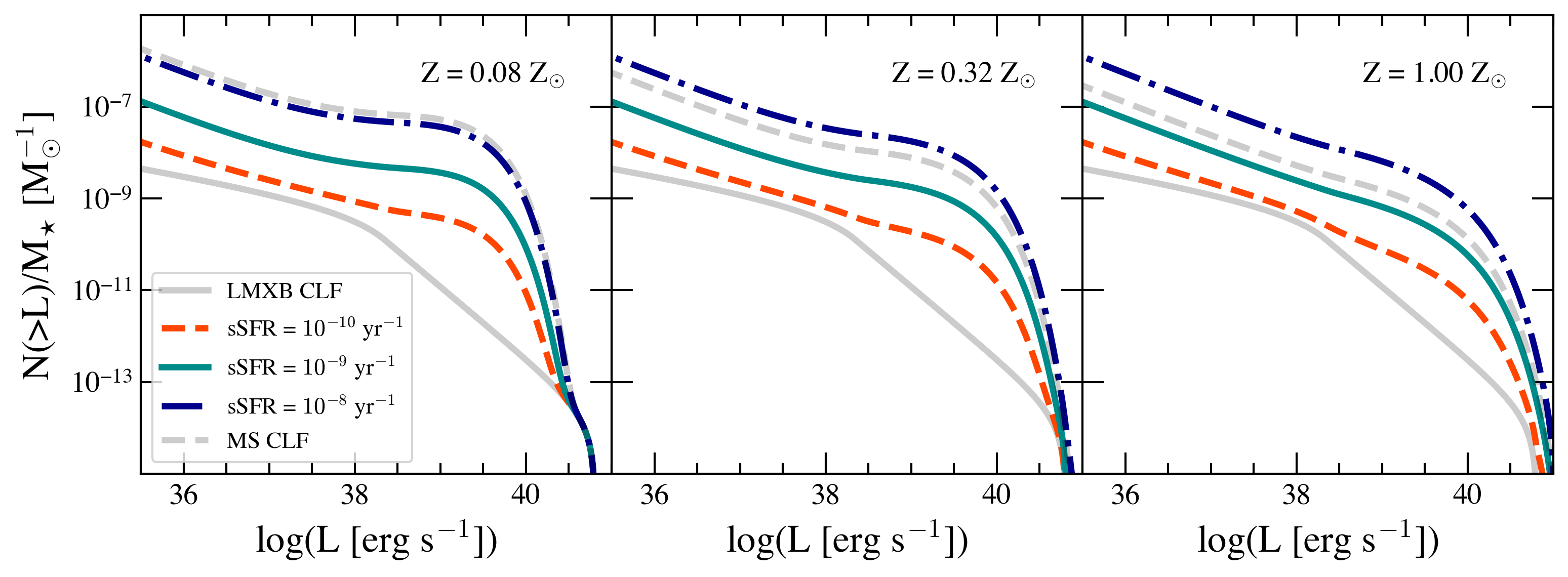}
\caption{Cumulative XLFs normalized by \mstar\ for varying sSFR ($\rm yr^{-1}$) at different metallicities ($Z$). The LMXB CLF is plotted in light grey in each panel for comparison. We also plot a typical CLF for a MS galaxy at each metallicity as the dashed grey line. For all metallicities, low sSFR curves more closely match the LMXB CLF due to the HMXB CLF being greatly dependent on SFR. Similarly, high sSFR causes the curve to be more aligned with the HMXB XLF. As $Z$ increases moving right, we observe changes in slope around transition points. The LMXB CLF component (derived from Equation \ref{eq:LMXLF}) is constant for each $Z$, indicating the changes in curve shape are due to the $Z$ dependence on the HMXB contribution (Equation \ref{eq:HMXLF}).
}
\label{fig:clf}
\end{figure*}

In these equations, all luminosities are taken to be in units of 10$^{38}$~ergs~s$^{-1}$. These XLFs predict the expected number of sources at a given luminosity for both HMXBs and LMXBs and their dependence on host stellar properties.  For a given choice of $M_\star$, SFR, and $Z$, we can calculate the total XLF and cumulative XLF (hereafter, CLF) following:
\begin{equation}\label{eq:xlf}
    \left.\frac{dN}{dL}\right \vert_{\rm tot} = \left.\frac{dN ({\rm SFR},Z)}{dL}\right \vert_{\rm HM} + \left.\frac{dN(M_\star)}{dL}\right \vert_{\rm LM},
\end{equation}
\begin{equation}\label{eq:clf}
     N(>L)_{\rm tot} = \int_\infty^L \left.\frac{dN}{dL}\right \vert_{\rm tot} dL.
\end{equation}
where the HMXB XLF only depends on SFR and $Z$ and the LMXB XLF only depends on \mstar. LMXB populations are commonly found in early-type, elliptical galaxies, which have a very small range of possible metallicities in the early universe. Observations of LMXBs are limited due to these host elliptical galaxies being devoid of nebular emission, which is necessary to establish any dependency on gas-phase metallicity \citep[e.g.,][]{fragos_x-ray_2013, fragos_energy_2013}. Therefore, we take the LMXB XLF to be independent of $Z$ for this work. Late-type, SF galaxies have an abundance of nebular emission due to stellar formation, which allows for much easier detections of gas-phase metallicity and a more obvious $Z$ dependence of the XLF. 

By summing the XLFs together and integrating to get the CLF, the three parameters that dictate the characteristics of the XRB population are \mstar, SFR, and $Z$. In Figure~\ref{fig:clf}, we display the CLFs, $N(>L)_{\rm tot}$ normalized by \mstar, for a variety of specific SFR (sSFR) and $Z$ values, where sSFR is defined as SFR/\mstar. We also plot the LMXB CLF that originates from Equation \ref{eq:LMXLF} and a typical CLF for a MS galaxy at each $Z$ for reference in each panel. As both $Z$ and sSFR increase, we observe a departure of the CLF curves from the basic LMXB CLF, displaying an increased abundance of HMXBs present in the populations. This trend is consistent with the dependency of the HMXB XLF (Equation \ref{eq:HMXLF}) on SFR and $Z$. Moving across the panels from left to right, we also notice smoother curves that trend away from the LMXB CLF. As sSFR and $Z$ increase, the CLF curves tend to deviate away from the LMXB CLF contribution (light grey), indicating a higher dependence on HMXB populations. 

Both Equation~\ref{eq:HMXLF} and \ref{eq:LMXLF} contain conditional statements where the slope of the XLF changes depending on the relation between the source's luminosity and the break luminosities ($L_{b,\rm HM}, L_{b,\rm LM}$). For HMXBs, the break luminosity represents the transition between different accretion mechanisms and marks whether low- or high-L sources will dominate the population. At $L \simlt L_{b,\rm HM}$, HMXBs are expected to be dominated by wind-fed systems and Be XRBs, which are fed by a decretion disk \citep[e.g.,][]{misra_x-ray_2023}. As $L$ surpasses $L_{b,\rm HM}$, an increasing fraction of the HMXBs are sustained through high-powered accretion from Roche-lobe overflow or very strong disk-fed accretion. For LMXBs, the XLF is expected to contain contributions from a diversity of donor-star and compact-object types, with the XLF break arising near the Eddington limit of the most massive neutron stars \citep[e.g.,][]{fragos_models_2008}.
Most LMXBs with $L < L_{b,\rm HM}$ are transient and have very low mass-transfer rates, which cause them to be quiescent and transient in nature. LMXBs with $L > L_{b,\rm LM}$ have mass-transfer rates high enough to sustain the accretion disk in order to stay stable and luminous \citep[][]{gilfanov_low-mass_2004}. 

This cutoff luminosities ($L_{c,\rm HMXB}, L_{c,\rm LMXB}$) allow for a population of very bright sources in all types of galaxies \citep[e.g.,][]{kovlakas2020census,i2022expanded}.
Sources with $L \simgt 10^{39}$~\lum\ are defined as ULXs, since their inferred isotropic luminosities exceed the Eddington rate of typical stellar mass BHs.  While the nature of ULXs is still uncertain and likely not homogeneous, several lines of evidence suggest that these systems are mainly compact objects, with some including NSs \citep[e.g.,][]{Bac2014}, that are experiencing mass transfer rates above the local Eddington limit \citep[e.g.,][]{kaaret_ultraluminous_2017}.  Models favor either a beamed geometry \citep[][]{KinA2023} and/or an alteration of the Eddington rate in the presence of strongly-magnetized NSs \citep[e.g.,][]{MusA2022}.


The CLF provides an estimate of how many sources are present and the luminosities at which they are most likely to occur. The number of sources observed in any given galaxy follows a Poisson distribution, so variations about the mean expected value are extremely common \citep[][]{gilfanov_statistical_2004}. This intrinsic stochastic scatter of the XLF causes wildly different predictions of the population's total luminosity ($L_{\rm X}$); (see \S \ref{subsec:strategy}). Stochastic variations in expected number of XRBs are observed to have relationships with SFR and $Z$, which affects the modeling of stellar feedback \citep[e.g.,][]{justham_another_2012, nikolic_2024}. Our aim is to generate realistic spectral models and this high stochasticity of the XRB power output must be addressed to achieve this goal. The presence of this stochasticity can cause scaling relationships from previous studies to fail to account for all characteristics of observed XRB emission \citep[e.g.,][]{persic_galactic_2007, mineo_x-ray_2012, basu-zych_evidence_2013, aird_x-rays_2018, aird_x-rays_2019, lehmer_x-ray_2019}. Poor sampling of the XLF will also cause these relationships to become non-linear and have antisymmetric distributions of uncertainty in the low-\mstar and low-SFR regime \citep[e.g.,][]{gilfanov_statistical_2004, justham_another_2012,lehmer_metallicity_2021}. \S \ref{sec:statistics} presents the statistical behaviors of the XLF and our sampling strategy to address this intrinsic stochastic nature.

\section{Statistical Properties of the XLF}\label{sec:statistics}

\subsection{The Expected Non-Linear $L_{\rm X}$-SFR Relation}\label{subsec:gilf}

\begin{figure}
\centering
\includegraphics[width=0.47\textwidth]{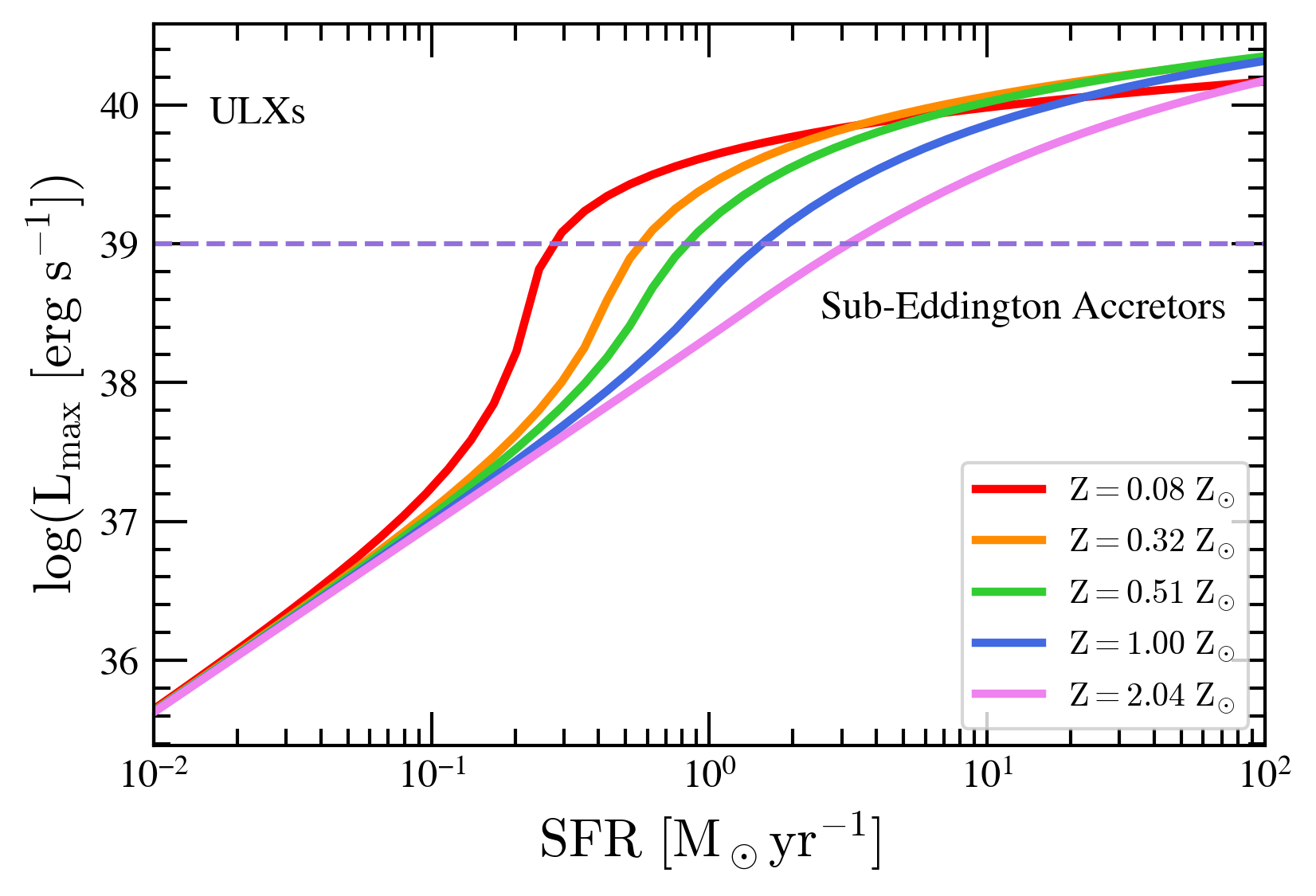}
\caption{The luminosity of the expected brightest source ($L_{\rm max}$) inferred from the HMXB XLF as a function of SFR for different metallicities. The purple dashed line denotes the dividing line between ULXs and more typical XRBs that are accreting at sub-Eddington rates. Since the HMXB population luminosity and spectrum is expected to be dominated by $L_{\rm max}$, ULXs play a significant role in shaping the integrated X-ray spectral output for galaxies with SFR~$\simgt$~0.2--2~\sfr.}
\label{fig:lmax_sfr}
\end{figure}

\citet{gilfanov_statistical_2004} demonstrated that the relatively shallow, high-$L$ slope of the HMXB XLF leads to a population-integrated luminosity ($L_{\rm X}$) that depends on the most luminous source within the population ($L = L_{\rm max}$): 
\begin{equation}\label{eqn:xlfint}
    L_{\rm X} \approx \int_0^{L_{\rm max}} L \frac{dN}{dL} \, dL,
\end{equation}
where $N(>L_{\rm max}) = 1$. $L_{\rm max}$ is intrinsic to the CLF curve and is defined as the luminosity of the brightest source expected by the CLF. $L_{\rm max}$ is plotted as a function of SFR for different metallicities in Figure \ref{fig:lmax_sfr}. We observe how the total luminosity will be dominated by ULXs or Sub-Eddington accretors depending on both SFR and $Z$. The type of source dominating $L_{\rm X}$ will also dominate the stochasticity in each population.

The HMXB XLF in Equation \ref{eq:HMXLF} has a slope $\gamma_1 = 1.74$ and $\gamma_2 < 2$ for typical metallicities, with an average slope $\langle \gamma \rangle \approx 1.6$ up to the cutoff ($L_{c,\rm HMXB}$). Below a certain SFR, the most luminous source will have $L_{\rm max} < L_{c, \rm HMXB}$. This implies: 
$$N(>L_{\rm max}) \propto {\rm SFR} \int^\infty_{L_{\rm max}} L^{-\langle \gamma \rangle} dL \propto {\rm SFR} L_{\rm max}^{-\langle \gamma \rangle + 1}$$
and setting $N(>L_{\rm max}) = 1$, further implies
$$L_{\rm max} \propto {\rm SFR}^{1/(\langle \gamma \rangle - 1)}.$$

In the low-$L_{\rm max}$ regime, the HMXB-population integrated luminosity is
$$L_{\rm max} \propto {\rm SFR}^{1/(\langle \gamma \rangle - 1)}.$$

since the average slope $\langle \gamma \rangle < 2$ for HMXB populations, and 
$$L_{\rm X} \propto {\rm SFR}^{1 - (\langle \gamma \rangle - 2)/(\langle \gamma \rangle - 1)},$$ 
which suggests $L_{\rm X} \propto {\rm SFR}^{1.67}$ for $\langle \gamma \rangle \approx 1.6$. 
However, in the high-SFR regime where the XLF is ``fully populated,'' meaning where the XLF curve closely matches the distribution of sources, $L_{\rm max} \rightarrow L_{c,\rm HMXB}$ and becomes independent of SFR, implying a linear relation $L_{\rm X} \propto \rm SFR$ in this regime. 

While we expect departures from linear luminosity relations in the low-SFR regime, galaxy XLFs are more complex than single, cutoff power-laws and also contain LMXB populations, which taken together depend on \mstar, SFR, and $Z$. Understanding how integrated X-ray luminosities, spectra, and their uncertainties vary with these properties is critical and requires a numerical sampling approach, which we describe in \S \ref{subsec:strategy}.


\begin{figure*}
\centering
\includegraphics[width=\textwidth]{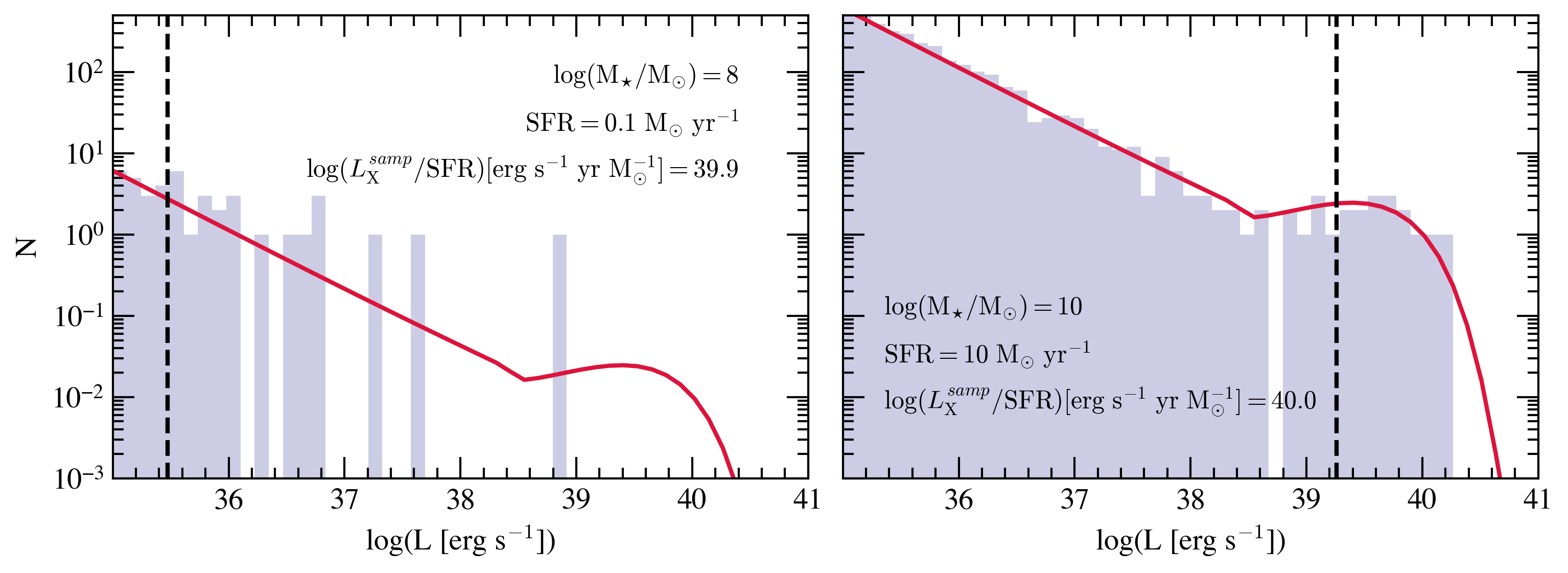}
\caption{Two XLFs for different populations with the same specific star formation rate (sSFR) of $ 10^{-9} \ {\rm yr}^{-1}$ and $Z = 0.30 \ Z_{\odot}$. These galaxies have the same sSFR, however, \mstar\ and SFR are able to vary quite a bit, causing the luminosity of the brightest source to take on a wide range of possible values. The continuous XLF is generated from the sum of Equation \ref{eq:HMXLF} and \ref{eq:LMXLF} and is shown in red. We also assigned luminosities to the sources present in these populations that are consistent with the XLF and their distribution is shown by the blue-shaded histograms. The vertical, dashed, black line gives the XLF's prediction of the log$L$ where the 15 brightest sources will have luminosities greater than or equal to this log$L$. We have found these sources to give an accurate representation of the whole population since they dominate the population's $L_{\rm X}^{samp}$. As \mstar\ and SFR increase, we see this point get farther to the right and eventually passes $L_{\rm b}$. To the right of this dividing line we observe the coarseness of the histogram compared to the relatively smooth nature at low $L$.}
\label{fig:XLF_hist}
\end{figure*}


\subsection{Sampling Strategy}\label{subsec:strategy}


Traditionally, sampling of populations from XLFs to obtain estimates of population-integrated $L_{\rm X}$ is performed statistically by (1) calculating from the CLF the total number of sources in the population that are more luminous than some minimum luminosity, $L_{\rm min}$ (i.e., $N(>L_{\rm min})$ in Equation~\ref{eq:clf}); (2) perturbing this value assuming a Poisson statistic with mean value set to $\mu = N(>L_{\rm min})$ to obtain $N_{\rm tot}$; (3) using the normalized CLF ($N(>L)$/$N(>L_{\rm min})$) as a statistical distribution to draw $L_i$ values for the $N_{\rm tot}$ simulated sources; and (4) performing the summation 
\begin{equation}\label{eq:sam}
    L_{\rm X}^{\rm samp} = \sum_i^{N_{\rm tot}} L_i
\end{equation}
to obtain a statistical draw of the population-integrated luminosity.

Figure~\ref{fig:XLF_hist} provides a histogram representation of the sampling of two XLFs with the same sSFR$ = 10^{-9}$~yr$^{-1}$ and $Z = 0.30 \ Z_\odot$: one XLF corresponds to a dwarf galaxy ($M_\star = 10^8$~\msol; SFR~=~0.1~\sfr) and the other to a more massive galaxy with properties comparable to some luminous infrared galaxies (LIRGs; $M_\star = 10^{10}$~\msol; SFR~=~10~\sfr). Both XLFs have the same shape according to Equations~\ref{eq:HMXLF}--\ref{eq:xlf} (red curves), but their normalizations differ by 2~dex. Full integration of these models via Equation~\ref{eqn:xlfint} yields $L_{\rm X}$/SFR~=~$1.7 \times 10^{39}$ erg s$^{-1}$ yr \msol$^{-1}$\ for both distributions. However, Figure~\ref{fig:XLF_hist} also shows that a sampling of these XLFs following the procedure above (shaded histograms) provide a highly stochastic estimate of $L_{\rm X}^{\rm samp}$, and we obtain $L_{\rm X}^{\rm samp}$/SFR~=~$7.9 \times 10^{39}$ erg s$^{-1}$ yr \msol$^{-1}$\ and $1.0 \times 10^{40}$ erg s$^{-1}$ yr \msol$^{-1}$\ for the dwarf and LIRG examples, respectively. The galaxy depicted in the right panel is an example of the XLF being fully-populated since the shaded histograms closely match the XLF curve.

\begin{figure*}
\centering
\includegraphics[width=0.9\textwidth]{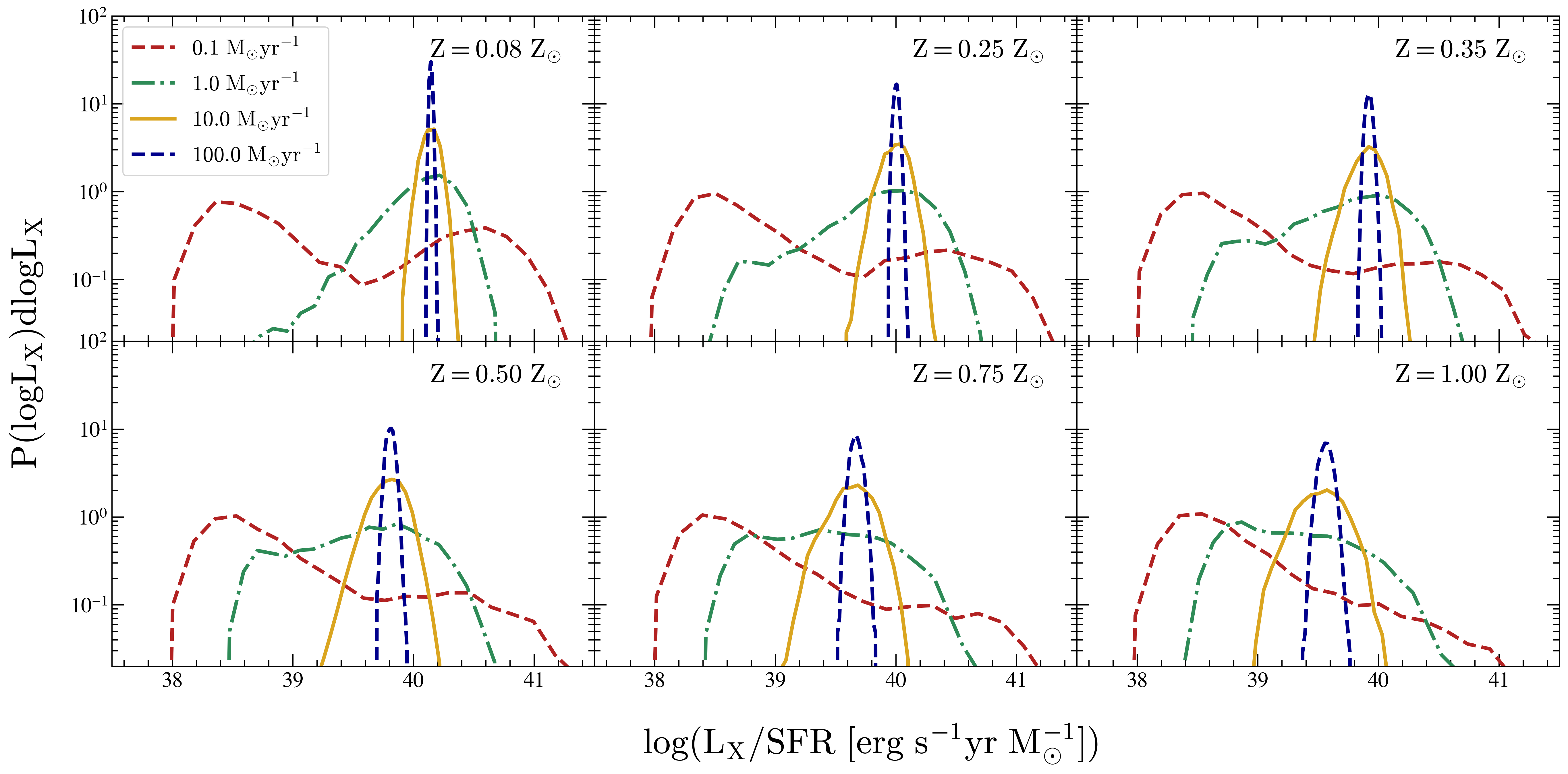}
\caption{Probability density functions (PDFs) of $\log (L_{\rm X}/\rm SFR)$ for HMXB populations to see the effects of SFR and $Z$ on the shape of the distributions. For all metallicities, we see a narrowing of the PDF with increasing SFR. The wide distributions at low SFR are results of sampling the low-L end of the XLF and higher degrees of fluctuation in their brightest sources (Figure \ref{fig:XLF_hist}).}
\label{fig:pdfs}
\end{figure*}

Looking at a given galaxy under stochastic variations, we refer to $L_{\rm X}^{\rm obs}$ as a measurement of the observed total luminosity of the galaxy. We can think of $L_{\rm X}^{\rm obs}$ as being one estimation of the random variable $L_{\rm X}^{\rm samp}$. Many simulations of estimating $L_{\rm X}^{\rm samp}$ will give a distribution of total X-ray luminosity, $L_{\rm X}$, for the XRB population. To assess whether such a measurement, $L_{\rm X}^{\rm obs}$, is consistent with a draw from the underlying XLF, we need to understand the distribution of $L_{\rm X}^{\rm samp}$, which will depend on \mstar, SFR, and $Z$.  While such distributions can be obtained using the above sampling procedure, repeated a large number of times, this method increases in computational cost with increasing SFR and $M_\star$, where luminosities of individual objects must be assigned for increasing numbers of simulated X-ray sources.

To mitigate the computational costs of simulating large numbers of sources, we took a hybrid approach.  Given that the brightest sources in the XLF distributions dominate the scatter of $L_{\rm X}^{\rm samp}$, we estimate $L_{\rm X}^{\rm samp}$ as:
\begin{equation}
    L_{\rm X}^{\rm samp} \approx \int_{\log L_{\rm min}}^{\log L(n)} \frac{dN}{d \log L } L \, d \log L + \sum_{i=1}^{n} L_i,
\end{equation}
where we adopt $\log L_{\rm min} ({\rm  erg~s^{-1}}) = 34$.  Here, the first term integrates the XLF function directly up until $n$ sources are expected above the luminosity $L(n)$, and the second term represents a sampled summation of the $n$ brightest sources above $L(n)$ following the methodology discussed above. Our estimation of $L_{\rm X}^{\rm samp}$ combines sampling the brightest sources and integrating the lower luminosity end of the XLF. We experimented with choices of $n$ for broad ranges of \mstar, SFR, and $Z$, and found that all probability distribution functions of $L_{\rm X}^{\rm samp}$ converged to common shapes for $n \simgt 15$. Therefore, we define $N~(>~L(n)) = n$, where $ n \sim $ Poisson($\lambda$), $\lambda = 15$ and thus $L(n)$ is represented in Figure \ref{fig:XLF_hist} by the dashed black line.

Our sampling strategy samples the $\approx~15$ (following a Poisson distribution where $ n \sim $ Poisson($\lambda$), $\lambda = 15$) brightest sources, regardless of whether they are HMXBs or LMXBs, and then integrates over the low-L end of the XLF curve. Since the population's $L_{\rm X}$ is greatly dominated by the most luminous sources, sampling of these sources is necessary to account for stochastic behavior. Comparing our sampling strategy to sampling every source expected by the XLF, we see a large improvement in computation time. For 5,000 sampling simulations of a Milky-Way-like galaxy with \mstar$ = 10^{10} $ \msol, $\rm SFR = 1$ \mdot, and $Z = 1 \ Z_{\odot}$, our method is $\approx~4.7$ times faster. Increasing the SFR to $100$ \mdot\ and thus causing the XLF to predict many more sources present, leads our method to become $\approx~304$ times faster than sampling every single source.

\begin{figure}
\centering
\includegraphics[width=0.47\textwidth]{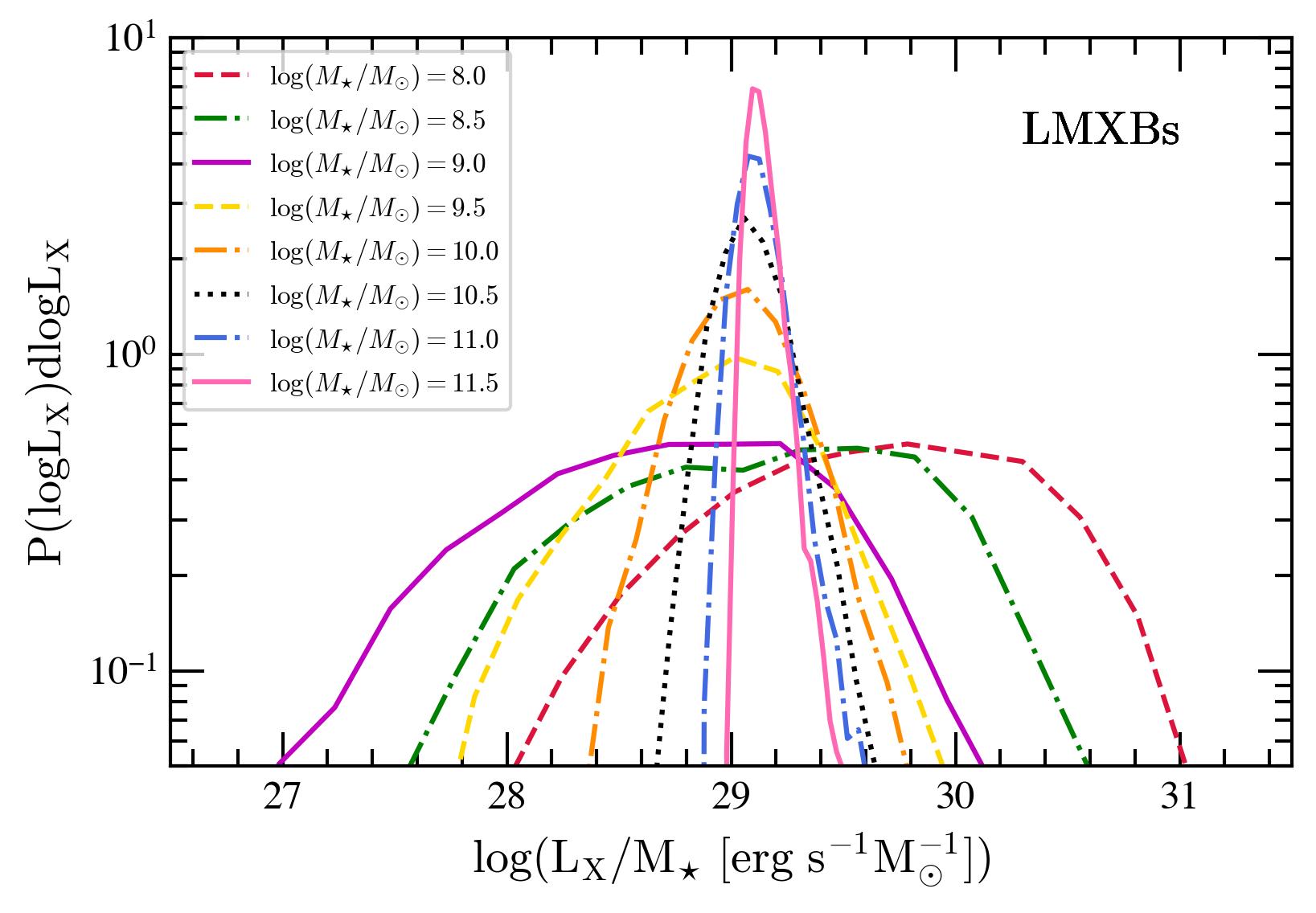}
\caption{PDFs of $\log L_{\rm X}$/\mstar\ for LMXB populations with varying stellar masses. These PDFs are independent of $Z$ since the LMXB XLF does not depend on $Z$ (Equation \ref{eq:LMXLF}). Similarly to the relationship between the PDF shape and SFR in Figure \ref{fig:pdfs}, we observe narrowing of the distributions as \mstar\ increases.}
\label{fig:pdf_lm}
\end{figure}




\begin{deluxetable*}{ccccc|cccc}
\tablenum{2}
\tablewidth{1.0\columnwidth}
\tablecaption{\label{tab:pdf_HM} Statistical Properties of the HMXB Probability Distributions}
\tablehead{
& \multicolumn{4}{c}{$Z = 1.0 \ Z_{\odot}$} & \multicolumn{4}{c}{$Z = 0.08 \ Z_{\odot}$} \\
& ($\rm SFR = 0.1$) & $(1.0)$ & $(10.0)$ & $(100.0)$ & ($\rm SFR = 0.1$) & $(1.0)$ & $(10.0)$ & $(100.0)$ 
}

%
%
\startdata
$\langle \log (L_X/\rm SFR) \rangle$ & $38.80$ & $39.29$ & $39.52$ & $39.56$ & $39.29$ & $40.05$ & $40.14$ & $40.14$ \\
$\tilde{\log (L_X/\rm SFR)}$ & $38.64$ & $39.24$ & $39.53$ & $39.56$
& $38.90$ & $40.10$ & $40.14$ & $40.14$ \\
$\sigma_{+}(\log (L_X/\rm SFR))$ & $+0.63$ & $+0.57$ & $+0.18$ & $+0.06$
 & $+1.63$ & $+0.24$ & $+0.07$ & $+0.01$ \\
$\sigma_{-}(\log (L_X/\rm SFR))$ & $-0.31$ & $-0.45$ & $-0.20$ & $-0.06$
 & $-0.53$ & $-0.32$ & $-0.08$ & $-0.01$ \\
\enddata
\tablecomments{
Mean $\log (L_X/\rm SFR)$, median $\log (L_X/\rm SFR)$, and upper and lower $1\sigma$ errors on the PDFs given in the $Z = 1.00$ and $0.08\ Z_{\odot}$ panels in Fig. \ref{fig:pdfs}. We find convergence between these values and the values reported by Table 3 in \cite{lehmer_metallicity_2021}.
}
\end{deluxetable*}

\begin{deluxetable*}{ccccccccc}
\tablenum{3}
\tablewidth{1.0\columnwidth}
\tablecaption{\label{tab:pdf_LM} Statistical Properties of the LMXB Probability Distributions}
\tablehead{
& ($\log M_\star = 8.0$) & $(8.5)$ & $(9.0)$ & $(9.5)$ & $(10.0)$ & $(10.5)$ & $(11.0)$ & $(11.5)$
}
\startdata
$\langle \log (L_X/M_\star) \rangle$ & $29.64$ & $29.15$ & $28.67$ & $28.92$ & $29.05$ & $29.10$ & $29.13$ & $29.14$ \\
$\tilde{\log (L_X/M_\star)}$ & $29.68$ & $29.19$ & $28.72$ & $28.95$ & $29.04$ & $29.08$ & $29.11$ & $29.12$ \\
$\sigma_{+}(\log (L_X/M_\star))$ & $+0.69$ & $+0.69$ & $+0.65$ & $+0.37$ & $+0.25$ & $+0.16$ & $+0.11$ & $+0.07$ \\
$\sigma_{-}(\log (L_X/M_\star))$ & $-0.76$ & $-0.82$ & $-0.80$ & $-0.47$ & $-0.24$ & $-0.14$ & $-0.09$ & $-0.05$ \\
\enddata
\tablecomments{Mean $\log (L_X/M_\star)$, median $\log (L_X/M_\star)$, and upper and lower $1\sigma$ errors on the PDFs given in Fig. \ref{fig:pdf_lm}. }
\end{deluxetable*}

Figure~\ref{fig:pdfs} displays the probability density functions (PDFs) of $L_{\rm X}^{\rm samp}$/SFR varying with SFR and $Z$ for HMXBs. We give probability distributions of $L_{\rm X}$ for $Z = 0.08, 0.25, 0.50, 1.00, 1.50,$ and $2.00 \ Z_{\odot}$ for SFR = $0.1, 1.0, 10.0,$ and $100.0$ \mdot. We find consistency between the $Z = 0.08$ and $1.0 \ Z_{\odot}$ panels of Figure~\ref{fig:pdfs} and the right panel of Figure~5 in \citet{lehmer_metallicity_2021}, which was based on stochastic sampling of the XLF across the full luminosity range used here.  Therefore, our efficient sampling method yields similar predictions of the $L_{\rm X}$-SFR-$Z$ relation that has been reported in previous studies \citep[e.g.,][]{basu-zych_evidence_2013,brorby_enhanced_2016}, while also reducing computation time. The statistical properties of the $Z = 0.08$ and $1.0 \ Z_{\odot}$ distributions are given in Table \ref{tab:pdf_HM}, which closely match statistical values presented in \citet{lehmer_metallicity_2021}. Similarly, Figure~\ref{fig:pdf_lm} shows the PDFs of $L_{\rm X}^{\rm samp}$/\mstar\ for LMXB populations with varying stellar masses and the statistical properties of these distributions are listed in Table \ref{tab:pdf_LM}. We observe closer values of the mean and median and more symmetric upper and lower confidence intervals for HMXB and LMXB populations with larger SFR and \mstar, respectively, since these distributions tend to approach a Gaussian nature. 

For HMXBs, we observe that regardless of metallicity, increasing SFR leads to smaller stochastic variations and approaches a symmetric, relatively-narrow Gaussian PDF, as expected from previous work \citep[see $\S$\ref{subsec:gilf};][]{gilfanov_statistical_2004}.  We see large spreads and skewness for low-SFR that is metallicity dependent. The middle panel shows $L_{\rm X}^{\rm samp}$ distributions for $Z = 0.08 \ Z_\odot$; the same metallicity as the populations shown in Figure~\ref{fig:XLF_hist}. The red curve at SFR $= 0.1$~\sfr\ shows two peaks in the $L_{\rm X}^{\rm samp}$ PDF.  

The high-$L_{\rm X}^{\rm samp}$ peak corresponds in simulations like the one shown in Figure~\ref{fig:XLF_hist}, where a small number of (or single) $L \simgt 3 \times 10^{38}$~\lum\ sources (possibly ULX(s)) are present and dominate the overall $L_{\rm X}^{\rm samp}$.  When such sources are not present the $L_{\rm X}^{\rm samp}$ tends to be significantly lower near the low-$L_{\rm X}^{\rm samp}$ peak in the PDF.  This phenomenon illustrates the large impact the brightest source will have on the expected total luminosity $L_{\rm X}^{\rm samp}$.

For LMXB populations (Figure~\ref{fig:pdf_lm}), we find somewhat different trends for the total luminosity PDFs.  While increasing \mstar\ narrows the PDF of $L_{\rm X}^{\rm samp}$, similar to the HMXB trends with SFR, the overall shape of the LMXB $L_{\rm X}^{\rm samp}$ PDF is mainly symmetric and relatively narrow over a broad range of $M_\star$.  These narrow distributions are expected due to the relatively steep $L \simgt 10^{38}$~\lum\ slope of the LMXB XLF, which leads to larger contributions from low-$L$ sources and less impact from the most luminous sources, compared to HMXBs. Since the XLF for LMXB contributions (Equation \ref{eq:LMXLF}) does not depend on metallicity, $Z$ is not a factor in these PDFs.


\section{Spectral Modeling}\label{sec:models}

\subsection{Luminosity Dependent Point-Source Spectral Models}\label{subsec:models}

To build X-ray spectral models for populations that are stochastically drawn from XLFs, we first constructed empirical distributions of X-ray point-source spectral shapes as a function of their observed luminosities in nearby galaxies.  To construct such empirical distributions, we made use of \chandra\ point-source catalogs from \citet{lehmer_empirical_2024} and the Physics at High Angular resolution in Nearby GalaxieS (PHANGS; Lehmer \etal\ in prep.), which respectively include 88 and 19 galaxies at $D \simlt 30$~Mpc.  The catalogs contain \chandra\ spectral extractions from {\tt ACIS Extract} \citep[{\tt AE}; ][]{Bro2010,Bro2012} for sources that span broad ranges of \mstar, SFR, and $Z$, which are plotted in Figure \ref{fig:mzr_msfr} (see Table B1 of \citet{lehmer_empirical_2024} for detailed observation information).  

Several of the galaxies in the \citet{lehmer_empirical_2024} sample subtend non-negligible solid angles on the sky, and thus large fractions of the point-sources are expected to be associated with background AGN, based on blank-field survey X-ray number counts contributions, $N(>L)_{\rm CXB}$.  To mitigate against significant background contamination in our distributions, we limited our analyses of the \citet{lehmer_empirical_2024} sources to those with high confidence of being intrinsic to their host galaxies, requiring that $P_{\rm bkg} \equiv N(>L)_{\rm obs}/N(>L)_{\rm CXB} < 0.1$, for a given point-source luminosity $L$.  \xray\ sources within the PHANGS galaxy sample have classifications that are based on a variety of multiwavelength properties (e.g., X-ray spectral shape, environmental locations in bulges or spiral arms, coincidence with star clusters, and mid-infrared colors that were consistent with belonging to their host galaxy and not background AGN).  For the PHANGS sample, we limited our analyses to sources that were classified as either ``HMXB'' or ``LMXB'' source types, which were identified as such using a succession of evidential indicators, including (1) associations with young and old star clusters or multiscale stellar associations (from \hst\ and \jwst\ SED fitting); (2) local environment ($\sim$100~pc) sSFR; and (3) lack of association with supernova remnants and background galaxies/AGN.  For both samples, we further limited our analysis to point-sources that had $> 10$ net counts in the 0.5--7~keV bandpass in order to ensure basic spectral shapes could be characterized. Diffuse hot gas emission is not included, as this analysis is restricted to resolved X-ray point sources and the stochastic behavior of the XLF. The contribution from hot gas is spatially extended and primarily affects the soft X-ray band ($\leq 2$ keV) \citep[][]{garofali2020x}.

\begin{figure*}
\begin{center}
\includegraphics[width=\textwidth]{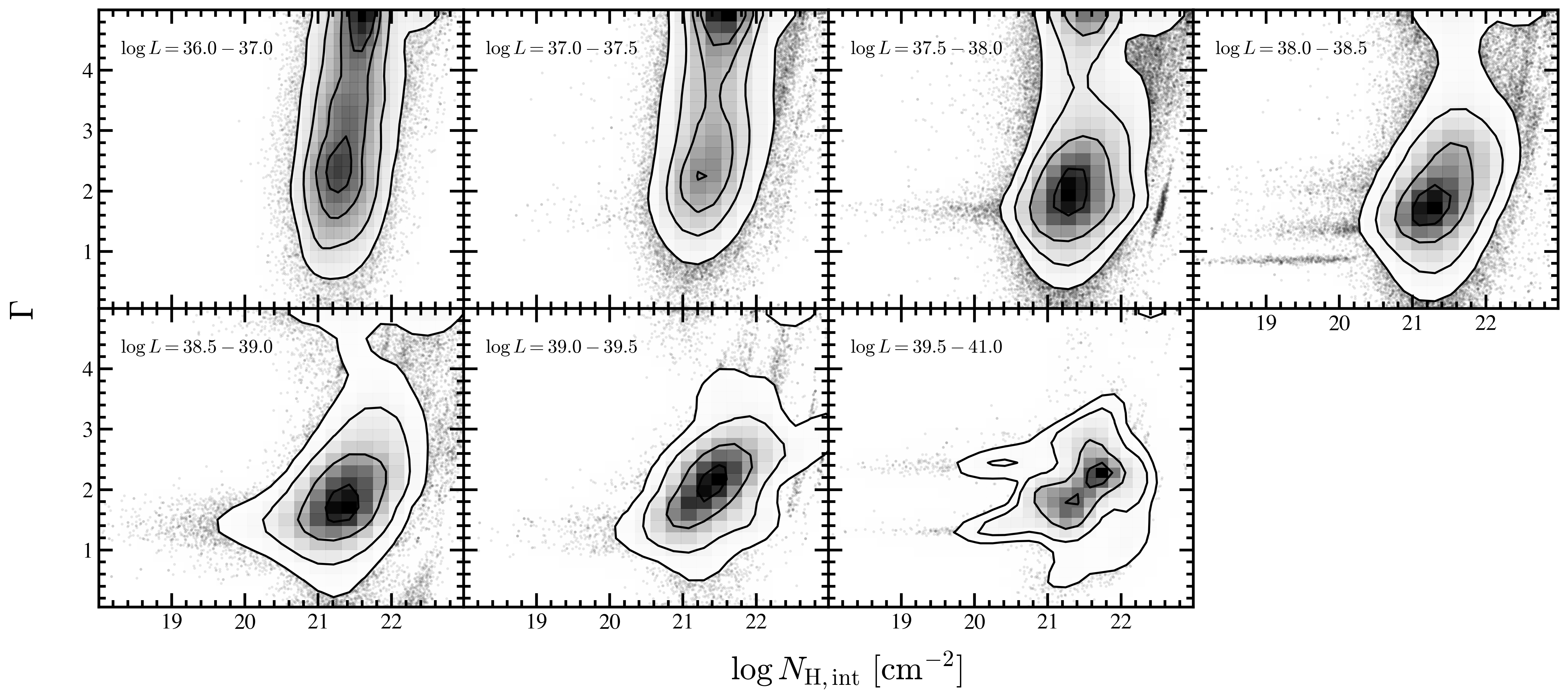}
\end{center}
\caption{Covariance contours of $\log N_{\rm H,int}$ and $\Gamma$ for several bins of $\log L$ created from sources in \citet{lehmer_empirical_2024} and PHANGS galaxies from Lehmer et al. (in prep.). We draw from these contours when generating spectral models to get realistic values for $N_{\rm H,int}$ and $\Gamma$ depending on the source's observed $\log L$.}
\label{fig:nhg}
\end{figure*}

In total, \nsrc\ X-ray point sources were identified with high confidence of being associated with their host galaxies ($P_{\rm bkg} \ll 0.1$ for most sources).  These sources span a luminosity range of $\log L \ ({\rm erg~s^{-1}}) =$~\hbox{36.7--40.2}.  For these sources, we performed X-ray spectral fitting, using the {\tt Sherpa} \citep[v4.17.1;][]{burke_sherpasherpa_2021} modeling package in {\tt python}, and \chandra\ data were extracted using {\tt CIAO} v15 with {\tt CALDB} v4.10.7.
We chose to adopt a forward-fitting approach, in which on-source data were modeled as the sum of a background model and an intrinsic source model.  The background model was determined using off-source background data, as extracted from {\tt AE}.  These background data were modeled using the {\tt cplinear} piecewise-linear model, which is appropriate for low-count background data.  The on-source data were modeled by fixing the rescaled background model and adding to it an intrinsic model representing the detected source spectrum.

For the intrinsic model, we adopted an absorbed power-law form that included both a fixed Galactic absorption component and a free variable intrinsic absorption component ({\tt tbabs$\times$tbabs$\times$pow} in {\tt XSPEC}).  The free parameters include the intrinsic column density, $N_{\rm H, int}$, the photon index, $\Gamma$, and the normalization on the power-law, $A$.  The Galactic absorption column, $N_{\rm H, Gal}$, for each source was fixed to the value appropriate for the sky location of each galaxy.   All fits were performed on unbinned spectral data spanning 0.5--7~keV using the Poisson $C$ statistic, as outlined in \citet{bonamente_2020}.  For sources with 0.5--7~keV counts $\simlt$100, we found that  $N_{\rm H, int}$ and $\Gamma$ were highly correlated and poorly constrained.  For these sources, we chose to adopt a loose Gaussian prior on $\log N_{\rm H, int} \sim  \mathcal{N}(21,0.3^2)$, as well as a uniform prior on $\Gamma \sim \mathcal{U}(-1,5)$.  We sampled the posteriors on these free parameters using Markov Chain Monte Carlo (MCMC) approach via the {\tt emcee} package \citep{foreman-mackey_2013}.

In Figure~\ref{fig:nhg}, we show the distribution of $\Gamma$ and $N_{\rm H, int}$ as a function of point-source luminosity $L$.  The shaded distributions in this diagram were constructed by smoothing over discrete points that represent the last 100 entries in the MCMC chains for each of the \nsrc\ sources. As such, the diagram therefore includes the effect of uncertainties on the fits of each of the sources. 


\subsection{Galaxy-Integrated X-ray Spectral Models}\label{subsec:sed}

To construct population-integrated X-ray spectral models, we sampled from the XLFs as described in $\S$\ref{subsec:strategy}, and assigned X-ray spectral models to each of the sampled XRBs by further sampling from the spectral-shape distributions that were constructed in $\S$\ref{subsec:models}.  Specifically, for a given XLF defined by the set of \mstar, SFR, and $Z$, we statistically drew individual luminosity values for the $\approx$15 most luminous sources and used the covariance distributions in Figure~\ref{fig:nhg} to statistically define $N_{\rm H, int}$ and $\Gamma$ values for each of the sources. The observed sources don't display a $\log L < 36.0$, but while sampling the CLF, it is definitely possible to find a source with low-L. Therefore, any sources we come across with a log$L < 36.0$ will draw $N_{\rm H,int}$ and $\Gamma$ from the $\log L = 36.0-37.0$ covariance contour in the first panel of Figure \ref{fig:nhg}. The final panel of Figure \ref{fig:nhg} corresponds to ULXs and is more sparsely populated than the others due to limited observational coverage. Additional ULX observations will be important for improving constraints on these parameters at high luminosities.

During the sampling process, the brightest sources are assigned luminosities which place each source into one of the bins of log$L$, which then corresponds to a covariance contour to assign values to $N_{\rm H,int}$ and $\Gamma$. These three parameters (expected log$L$, $N_{\rm H,int}$, and $\Gamma$) will be used as model inputs. Each source was modeled in {\tt XSPEC} using \hbox{{\ttfamily cflux(tbabs $\times$ pow)}}, where {\ttfamily cflux} is a convolution model applied directly to the power law with intrinsic absorption model, and calculates normalization of the overall model in flux units. 
The {\ttfamily cflux} model takes energy bounds and expected flux as parameters. We define an energy range of 0.5--8 keV to observe spectra over the full X-ray band and assign the statistically drawn luminosity to the flux parameter in this model component. 


%

\begin{figure}
\centering
\includegraphics[width=0.47\textwidth]{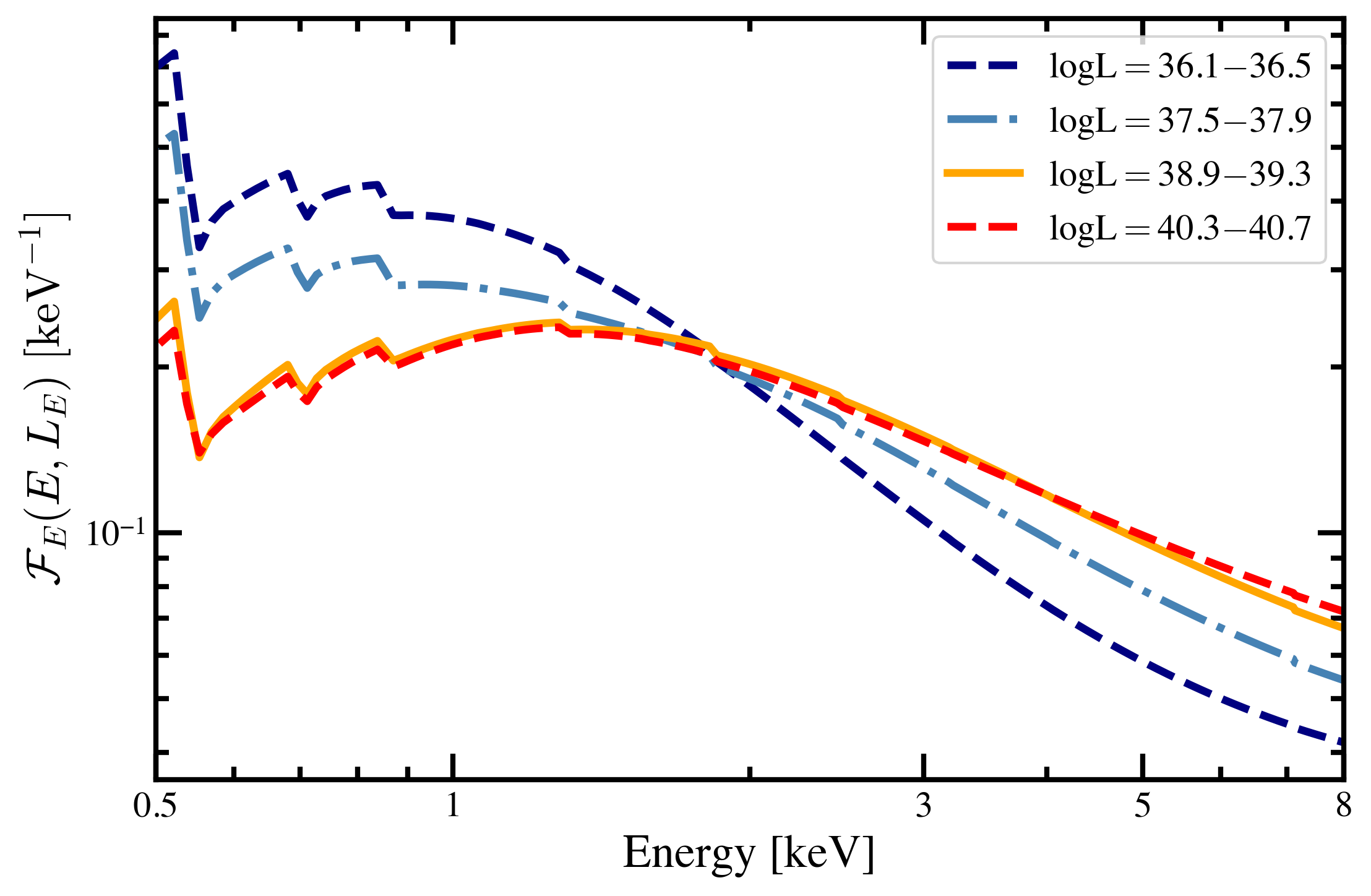}
\caption{Base functions, $\mathcal{F}_E(E, L)$, used in generating spectral models as a function of log$L$. Each function corresponds to a bin of $\log L$, which is denoted in the legend. These base functions are independent of the XLF and are used in Equation \ref{eq:spec} to construct our final spectral models. We observe the increase in the slope of the spectral models in the high-energy range.  We utilize the covariance contours of $\log N_{\rm H,int}$ and $\Gamma$ (Figure \ref{fig:nhg}) to construct these base functions for spectral modeling.}
\label{fig:speclogl}
\end{figure}

For any given galaxy, we use known characteristics of \mstar, SFR, and $Z$ and estimates of $L_{\rm X}$ resulting from our sampling procedure and calculations of total luminosity. As previously mentioned, the brightest sources in any given galaxy usually have the greatest influence on the galaxy's resulting spectra. We account for the lower $L$ end of the XLF by separating the luminosity range into bins and generating spectral models for each bin. By finding the spectral shape as a function of luminosity, we can scale these models based on the number of sources present in each bin for different galaxies. For older, elliptical galaxies, this method mitigates the chances for outlier low-L sources to dominate the overall spectra and yield an unreasonable slope. The resulting model for a galaxy is shown below:
\begin{equation}\label{eq:spec}
    L_E(E) = \int_{L_{\rm lo}}^{L_{\rm hi}} \frac{dN}{dL} L \,\mathcal{F}_E(E,L_E) \, dL + \sum_{i=1}^{n} L_i \, f_{E,i}(E,L_i),
\end{equation}
where, again, $ n \sim $ Poisson($\lambda$), $\lambda = 15$ and the values of $L_i$ and $f_i(L_i, E)$ are the values of the luminosity and unit-normalized spectral shape of the $i$th sampled source. We can also consider these base functions, $\mathcal{F}_E(E, L)$, that are used to generate the integrated spectral models (Figure \ref{fig:speclogl}) with weights applied that correspond to the number of sources expected by the XLF. The first term points to the low-L regime and accounts for all of the low luminosity sources. These sources don't impact the overall model much individually, which is why we integrate over this area. These sources must be accounted for to match what we see in observations, but sampling each source individually is unnecessary and adds extensive computation time. The second term is the sum of the individual spectral models for the $n$ brightest sources of the population. As previously mentioned, these sources dominate the resulting spectral model and we perform sampling on these sources to quantify the effect of stochastic scatter. Figure \ref{fig:speclogl} exhibits the difference in slopes of these models at the high energy end of the X-ray band with low $\log L$ corresponding to steeper, declining slopes. The change in spectral slope arises due to accretion flow becoming radiatively inefficient at low luminosities. In this regime, ions retain most of the energy and transfer less of it to electrons, which are responsible for producing the X-ray emission, leading to a different spectral shape. Absorption also becomes harder to constrain, which explains the large dip present in the low-energy band of all models, regardless of galaxy properties. Our resulting models exhibit a few dips due to absorption lines from various elements in the X-ray energy band. \citet{wilms_absorption_2000} discusses that the physical reasons for these drops at $\approx 0.54, 0.70,$ and $0.87$ are due to the absorption lines associated with O, the L transition of Fe, and Ne, respectively. These dips are present in essentially all resulting models we create due to the model's absorption component.

\begin{figure}[t]
\centering
\includegraphics[width=0.45\textwidth]{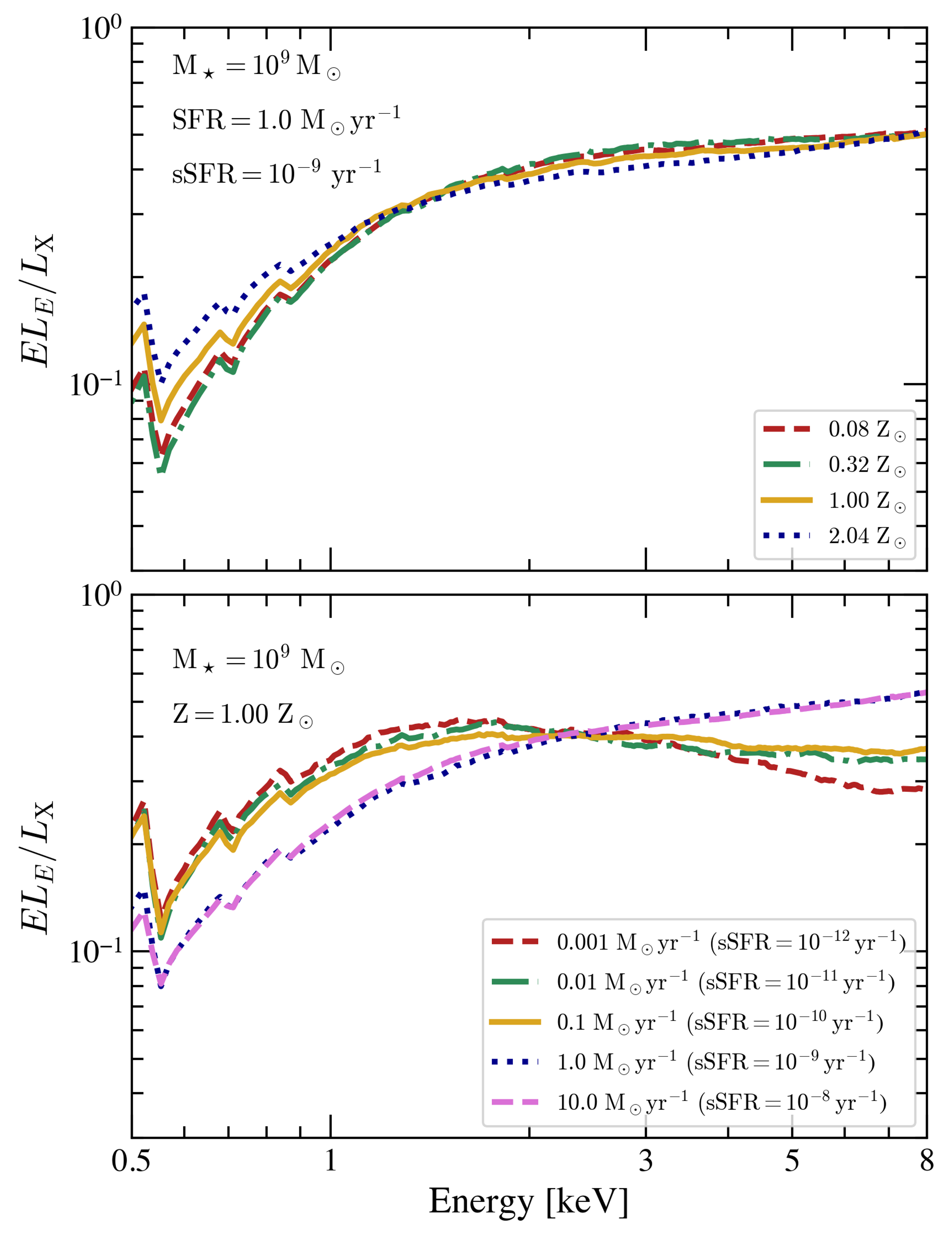}
\caption{Top panel: Spectra for a $10^9$ \msol\ galaxy at constant SFR $= 1 \ $\msol${\rm yr}^{-1}$ for varying $Z$. Bottom panel: Constant $Z = 1$ $Z_{\odot}$ for varying SFR. We isolate these variables to see the effect they each have on the resulting spectra. These models are normalized by the total integrated X-ray luminosity, $L_{\rm X}$.}
\label{fig:varyZSFR}
\end{figure}

We investigate the effects of changing $Z$ and SFR on spectral models of a constant \mstar\ in Figure \ref{fig:varyZSFR}. Changes in $Z$ while keeping SFR constant do not seem to impact the slope of the model, but that is expected due to the indirect dependence on Z in the equations defining the XLF (Equations \ref{eq:HMXLF} and \ref{eq:LMXLF}). If anyone notices any minor corrections that should be made before then, please let me know. We observe a very defined change in slope in the high-energy X-ray band as SFR is varied while $Z$ and \mstar\ are held constant. Increasing SFR leads to a flatter slope of the spectral model at high energies, indicating a significant contribution of hard X-ray emission, like Comptonization or non-thermal processes. For lower SFR, spectral emission peaks at lower energies, signaling that soft X-ray emission dominates in the form of thermal emission and disk accretion \citep[e.g.,][]{ebisawa1996spectral}. We observe an upturn in our base functions and models at high energies (see Figures \ref{fig:speclogl} and \ref{fig:varyZSFR}). In $EL_E$ space, a flat slope corresponds to where $\Gamma=2$. The covariance contours from which we draw our model parameters (Figure \ref{fig:nhg}) show that low luminosity sources often have $\Gamma > 2$, indicating a downward slope. The high luminosity bin ($\log L = 39.5-41.0$) is centered around $\Gamma = 2$, with fewer sources deviating away from this center. Because each model is constructed by summing contributions from multiple luminosity bins, the flatter spectra of the most luminous sources increasingly dominate at high energies, producing the observed upturn. We use \mstar-SFR and \mstar-$Z$ relations for MS galaxies to constrain both SFR and $Z$, respectively, given \mstar\ in order to narrow the area of uncertainty and produce realistic X-ray spectral models.

\begin{figure*}[h!tb]
\centering
\includegraphics[width=\textwidth]{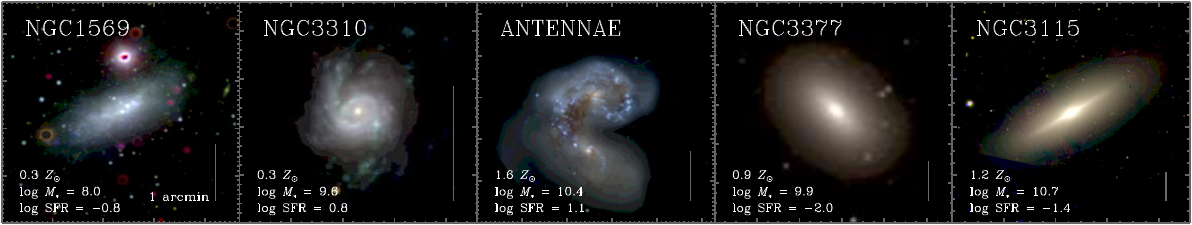}
\caption{False-color optical/near-IR images of the galaxies for which we generated spectral models (see $\S$\ref{subsec:gals} and Fig.~\ref{fig:5gals}) with \mstar, SFR, and $Z$ values annotated in each panel.  For reference of scale, a vertical bar of 1~arcmin length is displayed in the lower-right of each panel.  The displayed images were generated following the procedures in \citet{lehmer_empirical_2024}, and include red, green, blue imaging from SDSS ($g,r,i$; NGC3310, NGC3377), PanSTARRS ($g,z,y$; NGC1569), \hst\ and PanSTARRS (ACS F475W, $r,z$, NGC~3115), and \hst\ (ACS F435W, F550M, F814W; Antennae).  These galaxies were chosen to compare our spectral models with \chandra\ data across extremes of \mstar, SFR, and $Z$.}
\label{fig:photos} 
\end{figure*}

\begin{figure*}[h!tb]
\centering
\includegraphics[width=.99\textwidth]{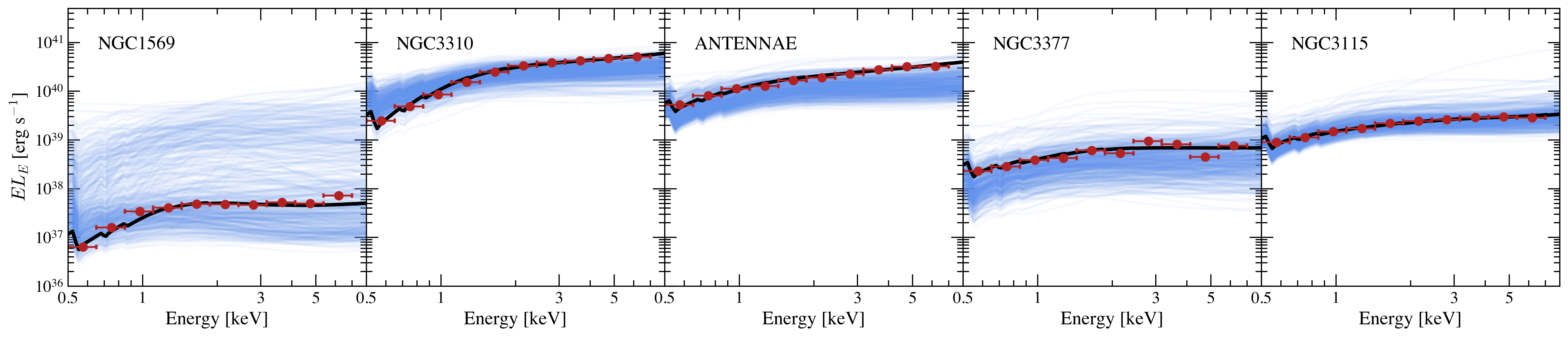}
\caption{\chandra\ data (red dots with errors) plotted with our simulated spectral models (blue) for the 5 galaxies shown in Figure \ref{fig:photos}, with the best-fit model of the 500 simulations denoted by the solid, black line. A higher concentration of models gives contrast between more probable regions and outliers. We observe a wider distribution of probable models for galaxies with low \mstar\ and low SFR, due to their higher degree of intrinsic stochastic scatter. }
\label{fig:5gals}
\end{figure*}

\section{Results}\label{sec:results}

\subsection{Model Testing of Local Galaxies}\label{subsec:gals}

We test our method on 5 galaxies covering a wide range of \mstar, SFR, and $Z$: NGC 1569, NGC 3310, the Antennae interacting galaxies (NGC 4038/4039), NGC 3377, and NGC 3115 (pictured in Figure \ref{fig:photos}). We developed spectral models for these galaxies and compared these models to \chandra\ data (Figure \ref{fig:5gals}). Our goal in selecting these galaxies specifically was to show how our models converge with spectral data for both late-type, star-forming galaxies and elliptical galaxies, regardless of SFR, \mstar, and $Z$. We plot 500 spectral models created using our sampling procedure and parameter assignment consistent with observations. Due to the statistical nature of this problem and our sampling procedure, multiple simulations are necessary to see which models constitute as reasonable versus outliers. For comparison, we plot 10-band \chandra\ photometric points on top of our models. The three late-type, star-forming galaxies in Figure \ref{fig:5gals} are NGC 1569, NGC 3310, and the Antennae interacting galaxies (NGC 4038/4039). We observe a large spread of model realizations for NGC 1569, which is due to its low SFR and low $Z$. As previously discussed (\S \ref{subsec:strategy}), low SFR and low $Z$ lead to wide distributions of expected population luminosity, yielding a larger range of realistic model results between simulations. NGC 3310 and the Antennae galaxies have higher SFRs, which tightens the space of possible spectral models. Each individual model is plotted with a low opacity to reveal areas and slopes of most probable models. Figure \ref{fig:5gals} also shows models generated for two early-type, elliptical galaxies: NGC 3377 and NGC 3115. These older galaxies have very low SFRs and a small number of bright sources. Their spectra are dominated by low-luminosity sources. We observe more outlier models for these galaxies due to the shallow slope of the XLF producing a broad distribution of $L_{\rm X}$ (Figure \ref{fig:pdfs}). Despite a few outliers, we find our spectral models provide a reasonable estimate of the spectral contributions in each region of the X-ray band. The solid black line in each panel of Figure \ref{fig:5gals} denotes the best-fit model out of the 500 simulated models for the \chandra\ data shown by the red points. 

We performed a likelihood analysis to quantify how well our models reproduce the \chandra\ observations. For each energy bin, we constructed a probability distribution using the distribution of model values generated by the 500 simulated samples and evaluated the probability corresponding to the observed \chandra\ measurement. The likelihood associated with a given observational data set, $\mathcal{L}_{\rm data}$, was calculated as the product of the probabilities across the 10 energy bins (indicated by the error bars on the red data points in Figure~\ref{fig:5gals}). To assess the significance of $\mathcal{L}_{\rm data}$, we also computed likelihood values, $\mathcal{L}_{i}^{\rm mod}$, for each individual simulated model, thereby constructing a distribution of model likelihoods. The null hypothesis probability, $p_{\rm null}$, was calculated as the fraction of simulations for which $\mathcal{L}_{i}^{\rm mod} < \mathcal{L}_{\rm data}$. 

The null hypothesis probability, $p_{\rm null}$, varies for our selected 5 galaxies in Figures \ref{fig:photos} and \ref{fig:5gals} with the smallest $p_{\rm null} = 0.079$ corresponding to the Antennae galaxies. We report $p_{\rm null} = 0.18$ and $p_{\rm null} = 0.26$ for NGC 3310 and NGC 3377, respectively. Our highest null hypothesis probabilities are $p_{\rm null} = 0.61$ and $p_{\rm null} = 0.81$ for NGC 1569 and NGC 3115, respectively. These results indicate that our simulated X-ray spectral models provide statistically acceptable reproductions of the observed \chandra\ spectra across both early- and late-type galaxies spanning a broad range of \mstar, SFR, and $Z$. While the distributions of models for NGC~3310 and the Antennae are less likely to reproduce the \chandra\ data, these cases are not statistical outliers.  The best simulated models for these galaxies (shown as bold black curves in Figure~\ref{fig:5gals}) provide very close reproductions of the observed \chandra\ spectra.
In future work, we plan to use SFH variations and better sample the most luminous sources (last panel of Figure \ref{fig:nhg}) to get a more accurate and representative X-ray spectral model. In particular, better sampling of the ULX population spectral properties will improve our models for high-mass SF galaxies like NGC~3310 and the Antennae.

We selected the example galaxies in Figure \ref{fig:5gals} to model galaxies with both low and high \mstar, SFR, and $Z$ in order to investigate the extensive applications of our spectral modeling procedure for different types of galaxies with high X-ray emissions. Table \ref{tab:demo} gives median spectral models for varying \mstar for constant $\rm SFR = 1.0$ \mdot and $Z = 1.0 \ Z_{\odot}$ and more models at varying SFRs and $Z$s is available for download (\url{https://doi.org/10.5281/zenodo.20126734}).

\subsection{The X-ray Spectra of Main Sequence Galaxies}\label{subsec:mainseq}

\citet{noeske_star_2007} discusses that a typical star-forming galaxy can only take on a range of realistic SFRs that are consistent with its \mstar\ and redshift ($z$) due to the physical processes behind star formation (SF) and those responsible for halting SF in the transition to quiescent galaxies. This relationship between \mstar, $z$, and SFR is referred to as the main sequence (MS) for star-forming galaxies. Previous work has indicated that star-forming galaxies constantly and moderately develop stellar populations as a result of the tight correlations between \mstar, $z$, and SFR on the MS 
\citep[e.g.,][]{rodighiero_multiwavelength_2014, ilbert_evolution_2015, schreiber_herschel_2015, aird_x-rays_2017}. Because the characteristics of MS galaxies follow tight constraints, we can estimate their SFR and $Z$ purely based on the galaxy's \mstar. 

\begin{figure}[hbt!]
\centering
\includegraphics[width=0.47\textwidth]{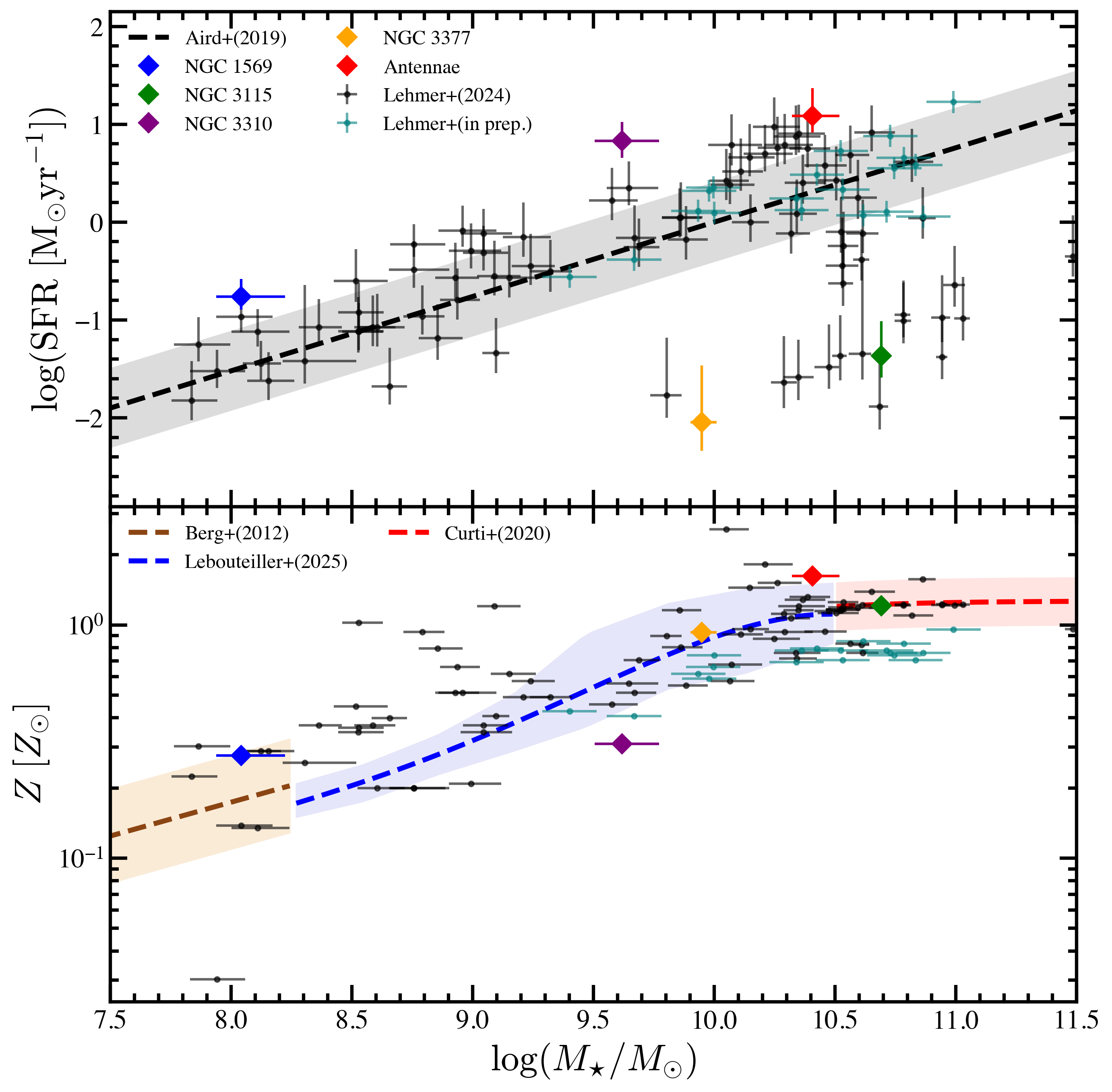}
\caption{Top panel: \mstar-SFR relation from \citet{aird_x-rays_2019}. Bottom panel: \mstar-$Z$ relations from \citet{berg2012direct,lebouteiller2025recovering} and \citet{curti2020mass}. The 5 galaxies shown in Figures \ref{fig:photos} and \ref{fig:5gals}, galaxies from \citet{lehmer_empirical_2024}, and PHANGS galaxies from Lehmer et al. (in prep.) are plotted in both panels.}
\label{fig:mzr_msfr}
\end{figure}

Prior research has found relationships between \mstar and SFR \citep[][]{aird_x-rays_2019} and \mstar and $Z$ \citep[][]{berg2012direct, lebouteiller2025recovering, curti2020mass} which narrow down the realistic values these three variables can take on in relation to each other. For the \mstar-$Z$ relation (MZR), we put together multiple relations to best match observations in different low-\mstar\ vs. high-\mstar\ regimes. We use the \citet{berg2012direct} MZR (Equation 14) for the low-\mstar\ regime $\log$(\mstar/\msol) $< 8.25$. The intermediate-\mstar\ range ($8.25 \leq \log$(\mstar/\msol) $ \leq 10.5$) uses the MZR from \citet{lebouteiller2025recovering} (Equation 10). Finally, for the high-\mstar\ regime ($\log$(\mstar/\msol) $> 10.5$), we use the MZR from \citet{curti2020mass} (Equation 2). Using these observationally-based relations together, we can predict a galaxy's SFR and $Z$ and generate an XLF given only its \mstar. These relations are plotted in Figure \ref{fig:mzr_msfr} with the \mstar-SFR relation in the top panel and the \mstar-$Z$ relations in the bottom panel. We also plot the 5 galaxies from Figures \ref{fig:photos} and \ref{fig:5gals} to show their relationship to what is expected for MS galaxies as well as galaxies from \citet{lehmer_empirical_2024} and the PHANGS galaxies in Lehmer et al. (in prep.) to show the spread of values these sources take on. We only use these relations to assign SFR and $Z$ to hypothetical MS galaxies for which these values are not already measured. All galaxies in Figure \ref{fig:mzr_msfr} are plotted to show their positions relative to the MS. These galaxies make up the covariance contours in Figure \ref{fig:nhg}, from which we draw our spectral model parameters. While we hope to add to this sample in the future, we make sure to include both MS galaxies and galaxies with more extreme values. We also recognize that the PHANGS galaxies (Lehmer et al. (in prep.)) are a bit below the \mstar-$Z$ relation, which may be due to differences in how the metallicities were calculated for that sample. The PHANGS metallicities come from \citet{groves2023phangs} using the S-cal method.



For MS galaxies, the \mstar-SFR relation has an uncertainty of $\pm 0.4$ dex \citep[][]{aird_x-rays_2019} and also depends on the redshift of our source, but for the purposes of this paper, we have set $z=0$ for our theoretical MS models. We adopt an uncertainty of $\pm 0.2$ dex for the \citet{berg2012direct} relation since scatter itself is uncertain as discussed in the text. We also adopt a $\pm 0.1$ dex uncertainty on the \citet{curti2020mass} relation, to be cautious around accounting for both scatter and parameter uncertainty. We calculated the scatter of the \mstar-$Z$ relation in \citet{lebouteiller2025recovering} for each bin of \mstar to obtain an accurate $1\sigma$ area of uncertainty. These uncertainties are all shown as shaded regions around the relations in Figure \ref{fig:mzr_msfr}.


\begin{figure}[hbt!]
\centering
\includegraphics[width=0.47\textwidth]{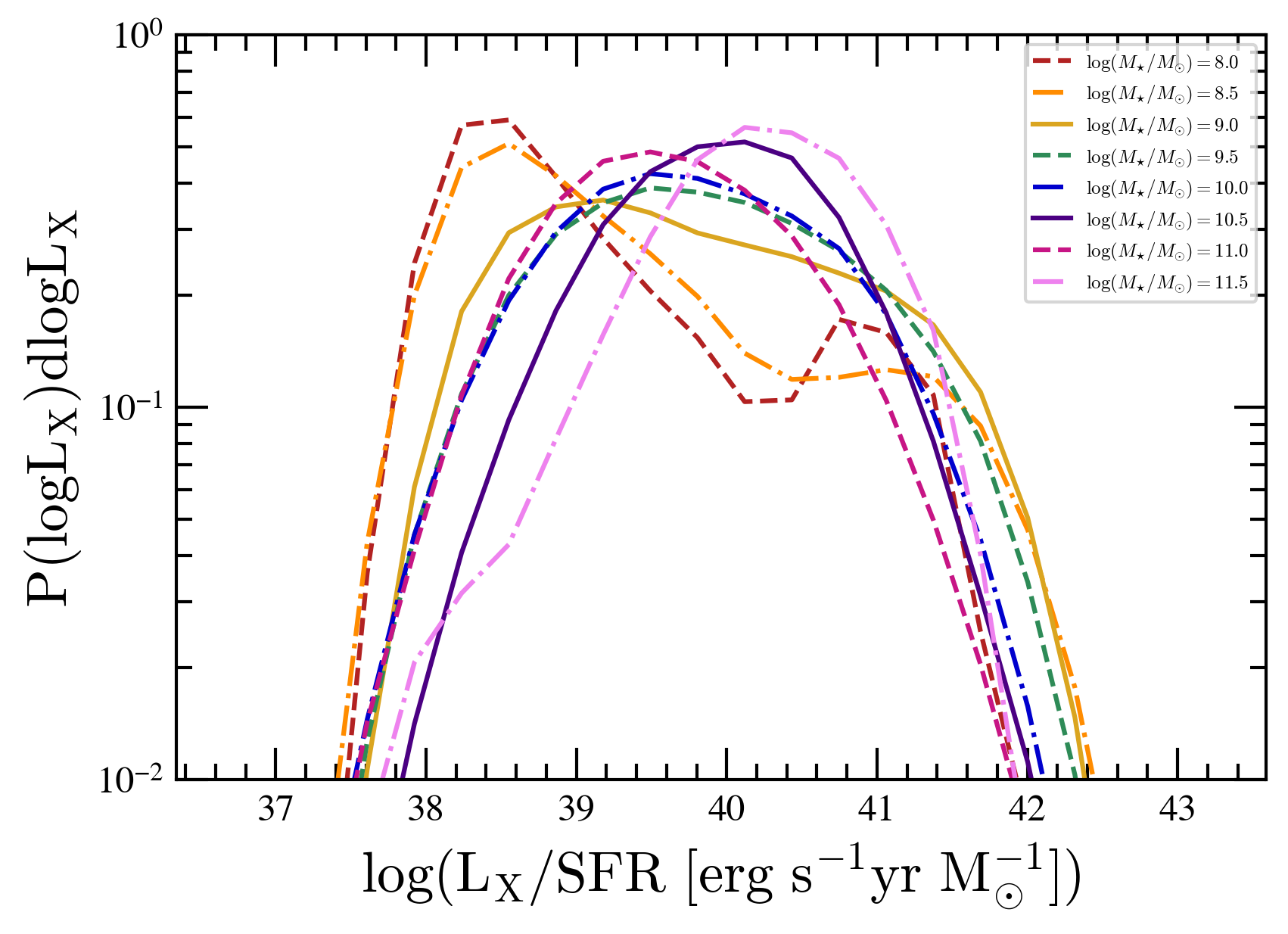}
\caption{Probability distributions of $L_{\rm X}$/SFR for XRB populations in MS galaxies with different \mstar. SFR and $Z$ are obtained through the \mstar-SFR and \mstar-$Z$ relations discussed in \S \ref{subsec:mainseq}.}
\label{fig:pdfs_M}
\end{figure}

We apply these relations and generate PDFs of $L_{\rm X}$/SFR for the entire population, including both HMXBs and LMXBs. Using the \mstar-SFR and \mstar-$Z$ relations discussed previously \citep[][]{aird_x-rays_2019,berg2012direct,lebouteiller2025recovering,curti2020mass}, we only give a value of \mstar\ and PDFs are output for a variety of populations (Figure \ref{fig:pdfs_M}). We observe similar trends as those in Figure \ref{fig:pdfs} where a higher SFR, and thus a higher \mstar, produces a Gaussian-like PDF with a lower influence of stochasticity. Conversely, the $L_{\rm X}$-SFR-$Z$ relation is much more subject to stochasticity for low-mass galaxies, which can be observed by the high levels of skewness and broad distribution of the low-\mstar\ PDFs. Low-mass galaxies show two areas of high probability due to the nonzero probability that they can host a high-luminosity source. As discussed in relation to Figure \ref{fig:XLF_hist}, the presence of this high-L source leads to a great deal of stochasticity in these low-mass XRB populations. The probability distributions of $L_{\rm X}$/SFR visually demonstrate how the sampling of the CLF curve greatly impacts the degree of stochasticity and uncertainty present in estimates of a population's $L_{\rm X}$. We find convergence between the results of our sampling method and previous work on the $L_{\rm X}$-SFR-$Z$ relation and using the \mstar-SFR and \mstar-$Z$ relations \citep[][]{aird_x-rays_2019, berg2012direct, lebouteiller2025recovering, curti2020mass}.

\begin{figure}[t]
\centering
\includegraphics[width=0.48\textwidth]{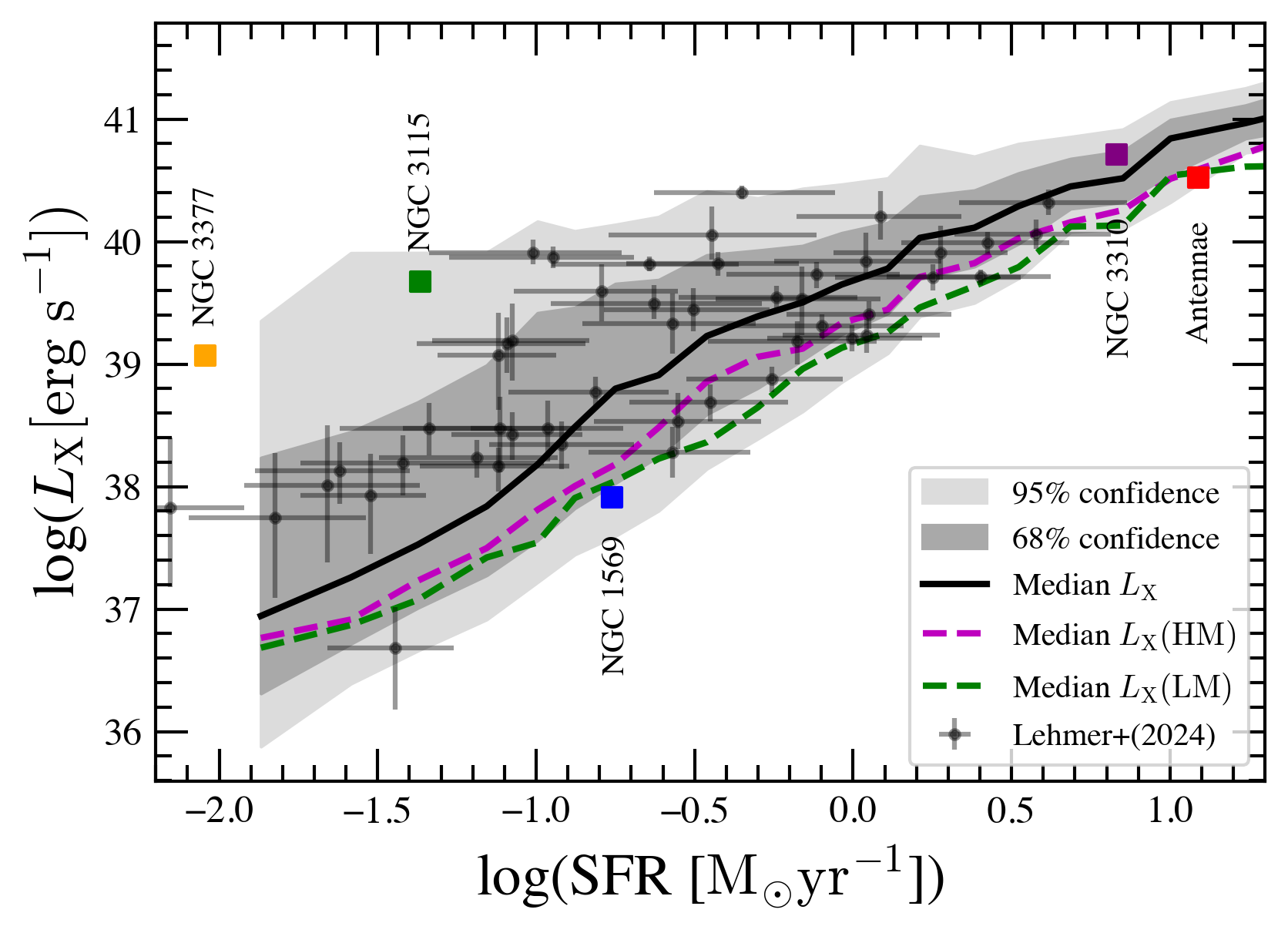}
\caption{Median $L_{\rm X}$ vs. SFR using the sampling procedure explained in \S \ref{subsec:strategy} for both types of XRBs (black solid line) and the median HM and LM components are shown with the dashed magenta and green lines, respectively. The 68\% and 95\% uncertainty are shown by the dark grey and light grey regions. Data points are galaxies from \citet{lehmer_empirical_2024} that are on the galactic main sequence. The galaxies modeled in Figure \ref{fig:5gals} are shown as colored squares with labels to indicate their relation to the MS. We see similarities between this our median trend and the left panel of Figure 5 in \citet{lehmer_metallicity_2021}.}
\label{fig:Lx-SFR}
\end{figure}

\begin{deluxetable*}{ccccccccccc}
\tablenum{4}
\tabletypesize{\footnotesize}
\tablewidth{1.0\columnwidth}
\tablecaption{\label{tab:demo} Spectral Model for SFR = $1$ \sfr; $Z = 1 Z_\odot$}
\tablehead{
%
 Energy & \multicolumn{10}{c}{$\log$($EL_E$) [erg s$^{-1}$]} \\
\vspace{-0.15in} \\
\cline{2-11}
\vspace{-0.15in} \\
(keV) & $(\log M_\star= 7.0)$ & $(7.5)$ & $(8.0)$ & $(8.5)$ & $(9.0)$ & $(9.5)$ & $(10.0)$ & $(10.5)$ & $(11.0)$ & $(11.5)$ 
}
%
\startdata
0.49 & $38.00$ & $38.19$ & $39.44$ & $39.74$ & $38.07$ & $38.17$ & $39.27$ & $38.96$ & $39.12$ & $39.65$ \\
0.51 & $38.02$ & $38.20$ & $39.44$ & $39.74$ & $38.15$ & $38.22$ & $39.29$ & $39.00$ & $39.15$ & $39.68$ \\
0.52 & $38.04$ & $38.22$ & $39.45$ & $39.74$ & $38.22$ & $38.27$ & $39.31$ & $39.04$ & $39.18$ & $39.71$ \\
0.54 & $37.94$ & $38.16$ & $39.41$ & $39.72$ & $37.99$ & $38.07$ & $39.17$ & $38.88$ & $39.10$ & $39.63$ \\
0.55 & $37.86$ & $38.12$ & $39.39$ & $39.70$ & $37.83$ & $37.94$ & $39.07$ & $38.77$ & $39.05$ & $39.57$ \\
%
\vdots & \vdots & \vdots & \vdots & \vdots & \vdots & \vdots & \vdots & \vdots & \vdots & \vdots \\
7.95 & $38.38$ & $38.77$ & $39.25$ & $39.29$ & $39.40$ & $38.65$ & $38.68$ & $39.53$ & $39.78$ & $40.47$ \\
7.96 & $38.38$ & $38.77$ & $39.25$ & $39.29$ & $39.40$ & $38.65$ & $38.68$ & $39.53$ & $39.78$ & $40.47$ \\
7.98 & $38.38$ & $38.78$ & $39.25$ & $39.29$ & $39.41$ & $38.65$ & $38.68$ & $39.53$ & $39.78$ & $40.47$ \\
7.99 & $38.38$ & $38.78$ & $39.25$ & $39.29$ & $39.41$ & $38.65$ & $38.68$ & $39.53$ & $39.78$ & $40.47$ \\
8.01 & $38.08$ & $38.08$ & $38.08$ & $38.08$ & $38.10$ & $38.15$ & $38.29$ & $38.86$ & $39.65$ & $40.30$ \\
\enddata
\tablecomments{One model simulation out of the 500 total models for 10 values of \mstar at $\rm SFR = 1$ \mdot\ and $Z = 1 \ Z_{\odot}$. A FITS file with all 500 models for the X-ray energy band ($0.5-8.0$ keV) and more combinations of \mstar, SFR, and $Z$ will be available for download. The complete tables holds models for all possible combinations of the following \mstar, SFR, and $Z$: $\log($\mstar/\msol$) = 7.0, 7.5, 8.0, 8.5, 9.0, 9.5, 10.0, 10.5, 11.0, 11.5$; $\rm SFR = 0.01, 0.1, 1.0, 10.0, 100.0$ \mdot; $Z = 0.08, 0.32, 0.51, 1.00, 2.04 \ Z_{\odot}$ $(\log(\rm O/H) +12 = 7.6, 8.2, 8.4, 8.69, 9.0)$.}
\end{deluxetable*}

Figure \ref{fig:Lx-SFR} depicts the relationship between median X-ray luminosity and SFR for galaxies on the main sequence with the black data points corresponding to the MS subset of the sample in \citet{lehmer_empirical_2024}. We observe convergence between our expected trend using our sampling method (light and dark grey confidence intervals) and where these points lie on the galactic main sequence. These trends and the shaded areas showing the 68\% and 95\% confidence intervals match Figure 6 of \citet{lehmer_metallicity_2021}.  

\begin{figure*}[h!tb]
\centering
\includegraphics[width=.8\textwidth]{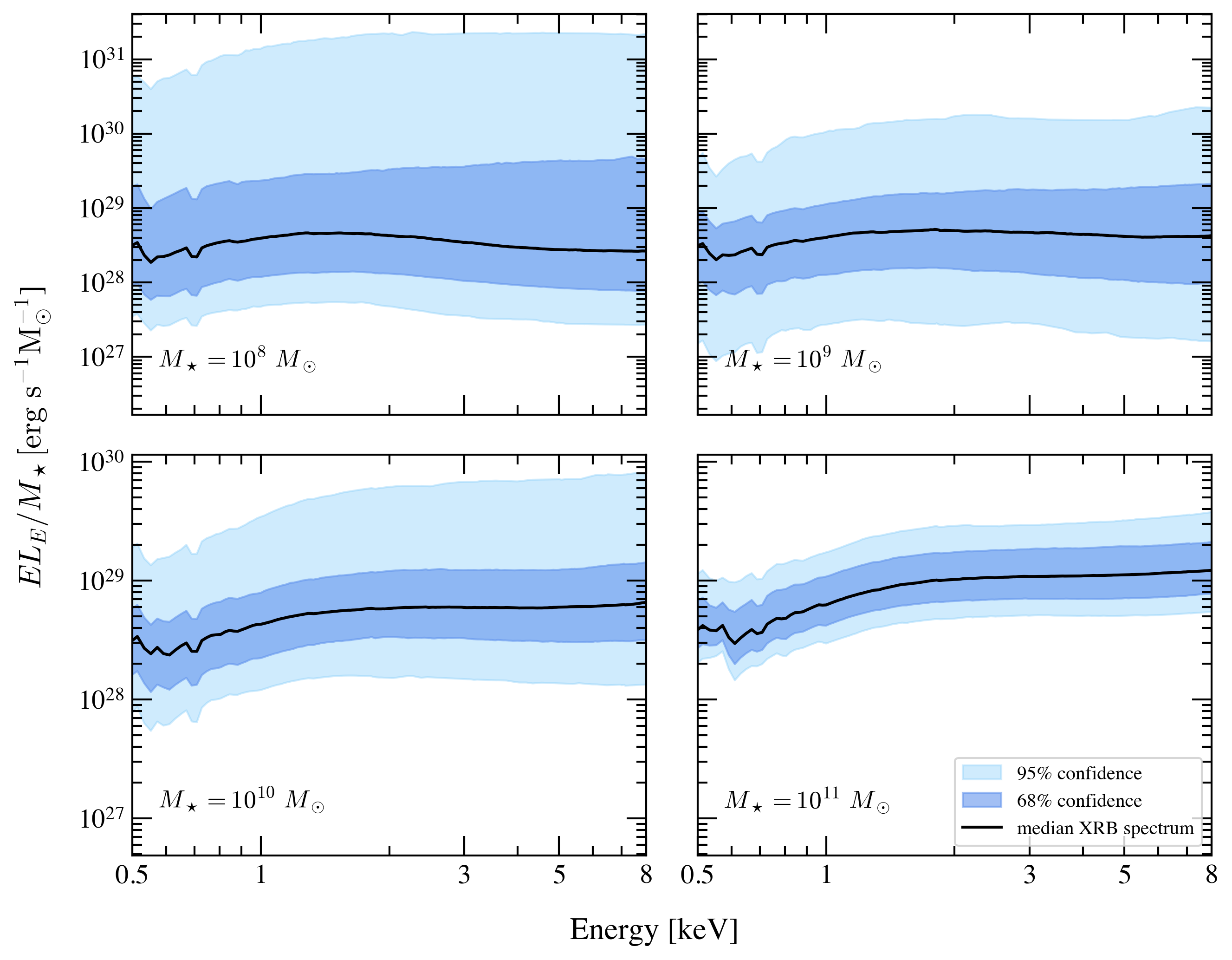}
\caption{Spectral models for typical MS galaxies with \mstar $ = 10^{8}, 10^{9}, 10^{10},$ and $10^{11}$ \msol. The median model is shown as a solid, black line, and the $1\sigma$ and $2\sigma$ confidence intervals are shown in dark and light blue, respectively. We notice larger areas of uncertainty at lower \mstar, which is consistent with the models shown in Figure \ref{fig:5gals}, since low \mstar\ corresponds to low SFR and low $Z$ following the \mstar-SFR and \mstar-$Z$ relations, and results in a high degree of stochasticity (Figures \ref{fig:XLF_hist}, \ref{fig:pdfs}, and \ref{fig:pdf_lm}).}
\label{fig:msmods}
\end{figure*}

Our sampling technique combined with these \mstar-SFR and \mstar-$Z$ relations allows us to input only a \mstar\ to return the galaxy's expected SFR, $Z$, a PDF of the population's total $L_{\rm X}$, and a predicted spectral model. This procedure accounts for stochastic scatter and reduced computational time to improve efficiency and accuracy. Figure \ref{fig:msmods} depicts spectral models for main sequence galaxies at a variety of \mstar's. Given a value of \mstar, the \mstar-SFR and \mstar-$Z$ will give reasonable estimates of SFR and $Z$ for many simulations. Figure \ref{fig:msmods} displays the median model (black), 68$\%$ (dark blue), and 95$\%$ confidence intervals (light blue). Similar to the galaxies shown in Figure \ref{fig:5gals}, we see a wider spread in low-\mstar\ galaxies since this typically corresponds to low SFR and low $Z$.

\section{Summary and Future Work}

This paper presents a procedure to generate realistic spectral models for galaxies hosting XRBs by accounting for statistical variations and intrinsic properties of the XLF. Key results are described below:

\label{sec:sum}

\begin{itemize}
    \item Stochastic scatter originates through the shallow slope of the XLF at low luminosities. Since the number of expected XRB sources follows a Poisson distribution, any small variation in number of sources constitutes a large possible range of $\log L$ when the XLF slope is shallow. This intrinsic stochastic scatter causes the estimation of the population's total X-ray luminosity, $L_{\rm X}$, to be complex.
    
    \item We account for stochasticity by sampling from the $\approx 15$ brightest sources and then integrating over the remainder of the XLF curve to get the probability distribution of $\log L_{\rm X}$. Galaxies with low SFR and low \mstar\ usually cannot support a high-L XRB source, but the probability is nonzero (Figure \ref{fig:XLF_hist}). Since the brightest sources dominate the $L_{\rm X}$ prediction, we conduct many simulations of our sampling process.
    
    \item As SFR and \mstar\ increase, probability distributions of $L_{\rm X}$ become narrower, more symmetric, and approach a Gaussian-like distribution showing that the degree of stochasticity decreases (Figure \ref{fig:pdfs}, Tables \ref{tab:pdf_HM} and \ref{tab:pdf_LM}).

    \item Using a combination of \chandra\ data and stacked spectra points, we have discovered differing degrees of covariance between $N_{\rm H,int}$ and $\Gamma$ depending on the source's luminosity (Figure \ref{fig:nhg}). We postulate that binning $\log L$ and drawing from these various contours will contribute to more realistic spectral models. 

    \item We have developed a procedure to generate spectral models of galaxies hosting XRBs by combining our sampling process and base functions (Figure \ref{fig:speclogl}) to build the final model (Equation \ref{eq:spec}). While the sampled sources will dominate the model's estimate of the total luminosity and degree of stochastic scatter, the base functions allow us to account for the low-L end of the population while saving computational time by not sampling every source expected to exist in each galaxy. 

    \item We find general convergence between our multiple simulated models and \chandra\ spectral data for the 5 galaxies shown in Figures \ref{fig:photos} and \ref{fig:5gals}. Although, our models for certain galaxies have a higher probability of fitting the data, we still see reasonable results. We selected these galaxies due to their locations relative to the galactic main sequence (Figure \ref{fig:Lx-SFR}). Despite these galaxies landing in different regions of the $L_{\rm X}$-SFR relation and the presence of stochastic variations, our models provide a reasonable estimation of spectral contributions for these galaxies with extreme values of \mstar, SFR, and $Z$ in the X-ray energy band (0.5-8 keV). 

    \item For galaxies on the MS, we use empirical parameterizations of the MS and \mstar-$Z$ relationship to determine reasonable values of SFR and $Z$ given a \mstar. We display these MS models in Figure \ref{fig:msmods} where we see a higher degree of stochasticity for low \mstar, consistent with the trends seen off the MS. 
    
\end{itemize}

For both typical galaxies on the MS and those having characteristics with extreme values, we are able to generate realistic spectral models with reasonable ranges of uncertainty using this method of statistical sampling. The statistical properties of the XLF have long been a source of uncertainty, but explicitly accounting for its stochastic nature allows for more accurate spectral models that are consistent with past observations of galaxies with high X-ray emissions. We have applied our knowledge of these common statistical variations to our sampling technique to model a handful of galaxies in different corners of the $L_X$-SFR relation plot (Figure \ref{fig:Lx-SFR}) as well as typical galaxies on the MS. Accounting for the intrinsic stochastic scatter of the XLF is critical to learning about the entire XRB population and creating a realistic spectral model in the X-ray band.

We will expand this work using new \chandra\ observations of the poorly studied low-Z regime to research how the previously mentioned relations evolve and affect spectral modeling for extremely metal poor galaxies (XMPs), as well as study ties between strong He II emission lines and presence of XRBs. The majority of early Universe galaxies are low-mass, highly deficient in metals, and high SFR. These galaxies are assumed to be main sources of cosmic reionization and are thought to be the primary contributors to the ionizing background at $z > 6$. XMPs are thought to dominate X-ray emission in the early Universe, meaning they play crucial roles in cosmic reionization, thermal evolution, and other related topics. We will focus on creating realistic spectral models for these objects since understanding their behavior and emissions are critical to comprehend other astrophysical phenomena. In this paper, we chose XLFs defined in \citet{lehmer_metallicity_2021} and 
\citet{lehmer_x-ray_2019} to study the effects of stochasticity for a galaxy's present day XLF. We eventually want to adapt our sampling process to study stellar age effects by using XLFs with star formation history (SFH) dependence \citep[e.g.,][]{gilbertson2022stellar, lehmer_empirical_2024}. We also plan to connect these X-ray spectral models with stellar population models of UV-to-IR data using spectral energy distribution (SED) fitting codes. We must understand the statistical significance and stochasticity present in special case outlier galaxies, MS galaxies, and XMPs to confidently and accurately incorporate this procedure into SED fitting codes. 

In future work, we can also further separate XRBs into four categories by using their compact object type and separation in NuSTAR X-ray color space \citep[][]{vulic2018black}, which cannot be done in \chandra\ energies alone. These four categories are BH binaries, magnetized and non-magnetized NS binaries, and ULXs. While power laws are useful for describing the shape of XRB spectral models in the X-ray energy band, they cannot be used to extrapolate to nearby energy ranges. We can determine physically-motivated models for each of these XRB source types and provide more detailed model fits to the point source emission that will also accurately extrapolate outside the X-ray band. Rather than relying on power laws and absorption models, we can model these XRBs by using different combinations of a simple blackbody model, accretion disk models with multiple blackbody components, thermal Comptonization models, and a self-irradiated funnel accretion model. By replacing our power law model with these different XRB model types, we can expect to significantly improve our fitting of \chandra\ data, along with allowing model fits to be extrapolated to both higher and lower energies. Defining parameters in these models will be motivated by intragalactic sources from \citet{vulic2018black} and thus will be physically motivated models. These models will also allow for the investigation of physical parameters such as disk and electron temperature, which is impossible with a power law. In the future, we hope to combine these more complex XRB models with the work done in this paper on accounting for stochastic scatter in X-ray spectral models. \\



We sincerely thank the anonymous reviewer for the time dedicated to reviewing this manuscript and for their helpful comments, which have improved the quality of this paper. We gratefully acknowledge financial support from Chandra X-ray Center grant AR2-23009X (E.G., B.D.L., I.P., Z.W.). C.R. acknowledges support from NASA RIA grant 24-RIA24-0038.


\facilities{{\it Chandra}, and some others}

\software{ {\ttfamily CIAO} \citep[v4.15]{fruscione_ciao_2006},  {\ttfamily DS9} \citep{joye_new_2003}, {\ttfamily Sherpa} \citep[v4.17.1]{burke_sherpasherpa_2021}, {\ttfamily XSPEC} \citep{arnaud_xspec_1996}}

\bibliography{stocscat.bib}{}
\bibliographystyle{aasjournal}



\end{document}